\documentclass[letterpaper]{JHEP3}

\newcommand{\Tr}{{\rm Tr}}

\newcommand{\tr}{\mathrm{Tr}}

\newcommand{\ba}{\begin{eqnarray}}
\newcommand{\ea}{\end{eqnarray}}
\newcommand{\nt}{\notag\\}
\newcommand{\nn}{\nonumber}
\newcommand{\back}{\!\!\!\!\!\!}

\newcommand{\bbZ}{{\mathbb Z}}

\usepackage{amsfonts}
\usepackage{amsmath}
\usepackage{amssymb}
\usepackage{cite}
\usepackage{comment}
\usepackage{graphicx}

\title{Moduli dynamics as a predictive tool for thermal maximally supersymmetric Yang-Mills at large $N$}

\author{Takeshi Morita  \\
 {\it Department of Physics,\\
Shizuoka University, 
836 Ohya, Suruga-ku, Shizuoka 422-8529, Japan} \\
{\it
Department of Physics and Astronomy\\
 University of Kentucky, Lexington, KY 40506, USA
 } \\
}

\author{Shotaro Shiba \\
{\it Maskawa Institute for Science and Culture, Kyoto Sangyo University\\
Kamigamo-Motoyama, Kita-ku, Kyoto 603-8555, Japan }\\
}

\author{Toby Wiseman\\
{\it Theoretical Physics Group, Blackett Laboratory, Imperial College, London SW7 2AZ, UK } \\
}

\author{Benjamin Withers\\
{\it Mathematical Sciences and STAG Research Centre, \\
University of Southampton, Highfield, Southampton SO17 1BJ, UK } \\
}

\preprint{MISC-2014-07}

\date{December 2014}

\abstract{
Maximally supersymmetric ($p+1$)-dimensional Yang-Mills theory at large $N$ and finite temperature, with possibly compact spatial directions, has a rich phase structure. Strongly coupled phases may have holographic descriptions as black branes in various string duality frames, or there may be no gravity dual. In this paper we provide tools in the gauge theory which give a simple and unified picture of the various strongly coupled phases, and transitions between them. Building on our previous work we consider the effective theory describing the moduli of the gauge theory, which can be computed precisely when it is weakly coupled far out on the Coulomb branch. Whilst for perturbation theory naive extrapolation from weak coupling to strong gives little information, for this moduli theory naive extrapolation from its weakly to its strongly coupled regime appears to encode a surprising amount of information about the various strongly coupled phases. We argue it encodes not only the parametric form of thermodynamic quantities for these strongly coupled phases, but also certain transcendental factors with a geometric origin, and allows one to deduce  transitions between the phases. We emphasise it also gives predictions for the behaviour of a large class of local operators in these phases.
}

\begin{document}
\setlength{\baselineskip}{18pt}

%-----------------------------------------------------------------
%
\section{Introduction}
%
%-----------------------------------------------------------------

The holographic correspondence \cite{Maldacena:1997re,Itzhaki:1998dd} is one of the most important discoveries in string theory.
It establishes two important lines of enquiry. Firstly it allows the analysis of strongly coupled phases of gauge theory through our understanding of semiclassical gravity, and secondly it allows investigation of the quantum aspects of gravity via direct computation in gauge theory.

Gravity analyses have revealed remarkable properties of supersymmetric gauge theories.
For example, the thermal energy density, $\epsilon$, of $(p+1)$-dimensional large-$N$ maximally supersymmetric Yang-Mills (SYM) theories ($p < 7$) at strong coupling ($1 \ll \lambda_p / T^{3-p} \ll N^{\frac{2(5-p)}{7-p}} $) are predicted by gauge/gravity duality to be precisely the same as that due to the semiclassical thermodynamics of a black D$p$-brane, and hence $\epsilon = a_p N^2 T^{\frac{2(7-p)}{5-p}} \lambda_p^{-\frac{3-p}{5-p}}$
where $\lambda_p$ is the 't Hooft coupling, and $a_p$ is a known constant. This behaviour is quite distinct both from the Stefan-Boltzmann law ($\epsilon \sim N^2T^{p+1}$) of weak coupling, or confining behaviour $\epsilon \sim O(N^0)$ typical of non-supersymmetric gauge theories.
Likewise, for the (2+1)- and (5+1)-dimensional superconformal field theories describing the low energy dynamics of M2 and M5 branes, holographic duality predicts from a black brane analysis that the thermal energy densities go as $\sim N^{3/2}T^3$ and $\sim N^{3}T^6$ respectively. 
Such gravity calculations have not only allowed  thermodynamic behaviour of these supersymmetric gauge theories to be precisely determined, but also have shown these behaviours are not those familiar in other gauge theory settings.

Ideally we would like to be able to compute these thermodynamic behaviours directly in the gauge theory. However, failing that, we would like to have a gauge theoretic understanding of how they arise. Unfortunately a naive extrapolation of weak coupling perturbation theory provides little insight into this, as clearly the perturbative behaviour $\epsilon \sim N^2T^{p+1}$ bears little resemblance to the behaviour above for black $p$-branes. Hope was glimpsed in the special conformal case of $p=3$, where weak and strong coupling do indeed have the same functional dependence on $N$ and $T$, differing only by a factor of $4/3$ \cite{Gubser:1996de, Aharony:1999ti}. However, as we see, this is a peculiarity of the conformal case, and in general the situation is more complicated. Indeed for the conformal ABJM theory of M2 branes weak coupling predicts $\epsilon \sim N^2 T^3$. Now the temperature dependence agrees with black M2-branes, due again to conformal symmetry, but the $N$ dependence is wrong.

The challenge then is to find new ways to compute thermal behaviour at strong coupling. 
One promising analytic approach, explored for $p=0$ is a method analogous to the mean field approximation \cite{Kabat:1999hp,Kabat:2000zv,Kabat:2001ve}, although recent work indicates that one still cannot see the above thermal behaviour \cite{Lin:2013jra}.
Numerical approaches were initiated for $p=0$ in \cite{Catterall:2007fp,Hanada:2007ti,Anagnostopoulos:2007fw,Catterall:2008yz} and confirmed the behaviour predicted by gravity, with the latest results being \cite{Kadoh:2012bg, Hanada:2013rga}. For $p=1$ numerical evidence \cite{Catterall:2010fx} also supports predictions from gravity. 
Other numerical approaches have focussed on dynamics within the thermal setting \cite{Asplund:2012tg}.
These numerical approaches are a very exciting prospect for the future, but are unlikely to shed light on \emph{why} the above behaviours emerge at strong coupling.
The purpose of this paper is to present a simple, yet predictive approach, that using direct calculation only in gauge theory allows one to understand the origin of the important features of the strong coupled thermal behaviour.

The approach we take builds on our recent work exploiting the special fact that the theories we are considering have exact moduli spaces associated to the Coulomb branch vacua. One may derive an effective moduli theory by integrating over the 
remaining degrees of freedom. This calculation is controlled by being far out on the Coulomb branch, implying these remaining degrees of freedom are very massive. Thus the moduli action can be computed as a weakly coupled loop expansion, although it is worth emphasising this is not the usual perturbative expansion in the 't Hooft or gauge coupling. 

The thermal strongly coupled regime does not correspond to these distant regions of the Coulomb branch, and so one must consider this moduli theory to be strongly coupled when describing the thermal physics of interest.
The crucial point is that whilst this moduli theory is computed in a weakly coupled regime, naive extrapolation to strong coupling appears to yield results that reflect the behaviour expected from gravity. It has long been known that it encodes the dynamics of a probe brane \cite{Banks:1996vh,Becker:1997xw,Okawa:1998pz, Kazama:2000ar,Buchbinder:2001ui,Baek:2008ws}, and this played a key role in the identification of M-theory and the holographic correspondence \cite{Maldacena:1997re,Itzhaki:1998dd}. Non-renormalisation theorems due to the supersymmetry play a crucial role in allowing these weakly coupled results to yield information about strong coupling \cite{Paban:1998ea,Paban:1998qy}.
Following earlier related ideas in the context of $p=0$ \cite{Horowitz:1997fr,Li:1997iz,Banks:1997tn,Li:1998ci} which considered a somewhat different phase than the one discussed above (associated to M-theory rather than D0-branes), and the work of \cite{Smilga:2008bt} again for $p=0$ which argued one could deduce the temperature dependence of black D0-brane behaviour, it was proposed in \cite{Wiseman:2013cda} that both the $N$ and temperature dependence of all the black Dp-branes could be recovered by naive estimates of the strongly coupled behaviour of the moduli theory. 
In fact the thermal behaviour of a certain mass deformation of the $p=0$ theory (the BMN matrix quantum mechanics \cite{Berenstein:2002jq}) was predicted in \cite{Wiseman:2013cda}, and this has been very recently confirmed in a dual supergravity analysis \cite{Costa:2014wya}.
Then \cite{Morita:2013wla} argued that using ABJM \cite{Aharony:2008ug} one could recover the thermal $N^{3/2}$ dependence of black M2-branes, and even, under modest assumptions about the M5-theory, a thermal $N^3$ behaviour in that case.

A new development was made in \cite{Morita:2013wfa}, where it was argued that for any black $p$-brane system, by constructing a field theory derived 
from considering the dynamics of nearly parallel branes weakly interacting gravitationally, then
naive thermal estimates at strong coupling
would yield the entire dependence on all physical parameters of the energy density, and further related the vevs of scalars in the field theory to the horizon radius of the black brane. 
Rather mysteriously it also allows certain transcendental factors with a geometric origin to be deduced. In the cases where the branes have a known gauge theoretic description (D$p$ and M2), this field theory was a subset of the full moduli theory, precisely due to the fact mentioned above that these moduli theories reproduce gravity dynamics.
More recently \cite{Morita:2014cfa} have applied these ideas to the D$1$-D$5$-P system to correctly reproduce the parametric dependence of the free energy on temperature, including certain transcendental factors. This generalisation to an intersecting brane system using the same methods of estimation in fact works despite important differences from the coincident brane case, and we regard it as good evidence for the validity of this approach.

In this paper we study the full moduli effective theory of maximally supersymmetric ($p+1$)-dimensional Yang-Mills at finite temperature, focusing on the bosonic sector, and also with a compact spatial direction. In addition to the phase described by a dual black D$p$-brane there are other regimes of the theory, some with black brane gravity duals related by S-duality or a lift to M-theory, and others without gravity dual \cite{Itzhaki:1998dd,Martinec:1998ja}. We also consider compactifying the theory on a spatial circle, which allows for phase transitions with dual Gregory-Laflamme descriptions \cite{Susskind:1997dr,Barbon:1998cr,Aharony:2004ig,Mandal:2009vz,Mandal:2011hb}. Using the ideas of \cite{Morita:2013wfa}, we argue that the moduli theory encodes all the strongly coupled phases, and also predicts where transitions will occur. 
Furthermore it allows not only the behaviour of the energy density to be found, but also predicts the behaviour of wide classes of local operators involving the scalars and gauge fields and their derivatives. Such predictions are highly non-trivial from the dual perspective, and we expect may be tested where gravitational duals exist.
In addition, the moduli effective theory reproduces not only the 
dynamics of the dual D$p$-branes as discussed in \cite{Morita:2013wfa}, but also encodes correctly the dynamics of the M2, M5, F1 and NS5 branes that are dual in certain parameter regimes.

This paper is arranged as follows. We begin in Section~\ref{sec:SYM} by discussing the effective moduli approach to study the strongly coupled thermal phases of SYM theory. This section considers the large $N$ regime where the temperature is of order one in units of the 't Hooft coupling - this is the regime where the black D$p$-brane is the gravity dual. In particular we emphasise that this moduli approach predicts the behaviour of the thermal vevs of large classes of local operators.
Then in Section~\ref{sec:circle} we show how the moduli theory is modified when the SYM has a spatial direction that is compact. We argue that the moduli theory then correctly predicts a transition in thermal behaviour which is dual to the gravity Gregory-Laflamme phase transition.
In Section~\ref{sec:corrections} we discuss the corrections to the moduli theory that become relevant at very strong coupling, where the temperature in units of the 't Hooft coupling scales with a power of $N$. We argue that the various transitions in behaviour predicted by gravity may be understood directly in  the moduli theory.
The main result of these discussions is that the strongly coupled phases of the SYM phase diagrams, with a compact spatial direction, can be elucidated simply by the moduli theory. In contrast the dual description is very complicated involving many different dual gravity theories, branes and transitions. This entire picture is discussed in detail in the Appendix~\ref{app:fullstructure}. 
We conclude the paper with a discussion in Section~\ref{sec:outlook}.

%-----------------------------------------------------------------
%
\section{Review of effective moduli theory for thermal maximally supersymmetric Yang-Mills}
\label{sec:SYM}
%
%-----------------------------------------------------------------

We review the derivation of the moduli theories for $(p+1)$-dimensional maximally supersymmetric Yang-Mills theories. We then review how this theory encodes 
the dynamics of weakly interacting D$p$-branes 
and how a naive extrapolation to strong coupling yields the thermal behaviour of black D$p$-branes 
of the holographic gravity dual
\cite{Morita:2013wfa}.
The $(p+1)$-dimensional maximally supersymmetric $U(N)$ Yang-Mills theory has an action,
\begin{eqnarray}
{S}_{\text{SYM}} = \frac{N}{\lambda_p}  \int dt d^px \, \tr\left[  - \frac{1}{4} \mathbf{F}_{\mu\nu}^2 - \frac{1}{2} D^\mu \mathbf{\Phi}^I D_\mu \mathbf{\Phi}^I + \frac{1}{4} \left[ \mathbf{\Phi}^I , \mathbf{\Phi}^J \right]^2  \right] + \mathrm{fermions}
\label{eq:SYM}
\end{eqnarray}
where the scalar fields $\mathbf{\Phi}^I$ ($I = 1, 2,  \ldots , 9-p$) and gauge field $\mathbf{A}^\mu$ ($\mu = 0, \ldots,p$) are $N \times N$ hermitian matrices transforming in the adjoint of the gauge group. 
$\lambda_p$ is the 't Hooft coupling defined as 
\begin{align}
\lambda_p \equiv g_{YM}^2 N 
= (2\pi)^{p-2} N g_s (\alpha')^{\frac{p-3}{2}} 
\end{align}
where $g_s$ and $\alpha'$ are the string coupling and the Regge slope respectively.

Consider the thermodynamics of this theory on flat space and finite temperature $T$. Then the theory has two dimensionless parameters: the dimensionless effective coupling 
\begin{align}
\lambda_{eff} = \lambda_p \, T^{p-3} \, , 
\end{align}
and $N$.
We focus on the large-$N$ limit where thermodynamical properties  are  determined by the strength of this effective coupling $\lambda_{eff}$.
 We note that for $p=3$ the theory is famously conformal. For $p>3$ the theory becomes subtle due to its non-renormalizability.
 However, various arguments in string theory indicate that the theory may have a good UV completion and hence can be thought of as an effective low energy description of a more fundamental theory on the world volume of D$p$-branes \cite{Itzhaki:1998dd}. 
\footnote{
We note that as argued in \cite{Itzhaki:1998dd,Peet:1998wn} the gravity analyses show that  the canonical ensemble may not be well defined for $p \ge 5$.
In these cases a natural dimensionless parameter is $\lambda_p \chi^{p-3}$ rather than $\lambda_{eff}$, where $\chi$ is the characteristic scale of the adjoint scalars.
}

\subsection{The perturbative limit}
\label{sec:weak}

In the weak coupling region ($\lambda_{eff} \ll 1$) one may perform the standard QFT weak coupling perturbation around $ \mathbf{\Phi}^I =0$.
Through a dimensional analysis, we can estimate the thermal energy density $\epsilon$ and the scale of the scalar fields as\footnote{
\label{fn-weak}
These dimensional arguments are naive as we have ignored IR divergences.
For $p=0$,  $\epsilon$ and $\Phi$ diverge at zero coupling because of the large fluctuations of the thermal zero modes of the scalars.
However the commutator square interactions suppress this divergence non-perturbatively and then they scale as $\epsilon \sim N^2 T$ and $\Phi \sim (\lambda_0 T)^{1/4}$ for small $\lambda_{eff}$.
Indeed the scaling for the scalar differs from the naive dimensional analysis (\ref{eq:SYM-weak}).
 See \cite{Hotta:1998en, Mandal:2009vz} for the details.  
}
\begin{align}
\epsilon \sim N^2 T^{p+1}, \qquad
 \Phi  \sim \sqrt{ \lambda_{eff} } \, T.
 \label{eq:SYM-weak}
\end{align}
Here we see the Stefan-Boltzmann law mentioned above for the energy density. The coupling dependence of the scalar is obtained by renormalizing the scalar to the canonical one, and the $N^2$ is determined by the number of UV degrees of freedom.
As emphasised in the introduction, a naive extrapolation of these formulae into the regime where $\lambda_{eff} \gg 1$ fails to reproduce the behaviour of the theory that we deduce from holographic arguments.

\subsection{Moduli dynamics from a weakly coupled regime on the Coulomb branch}

At zero temperature the maximally supersymmetric theory admits an exact moduli space, parameterized by the scalar field $ \mathbf{\Phi}^I$ hermitian matrices being constant and commuting. We may write such vacua as,
\begin{eqnarray}
(\mathbf {A}^\mu)_{ab} = a^\mu_a \delta_{ab} \; , \quad  ( \mathbf{\Phi}^I )_{ab} = \phi^I_a \delta_{ab}\; , \quad \mathbf{\Psi}_{ab}^\alpha =0
\end{eqnarray}
where $\phi^I_a$ are constants parameterizing the breaking of $U(N) \to U(1)^N$, and we have also included constant commuting gauge fields parameterized by the constants $a^\mu$. Such a vacuum configuration encodes a mass scale, $| \vec{\phi}_{ab}|$, due to the separation of the scalars where we have introduced the vector notation $\phi^I_a \leftrightarrow \vec{\phi}_a$, thinking of the moduli as being valued as vectors in $\mathbb{R}^{9-p}$. We have further introduced the notation $\vec{\phi}_{ab} = \vec{\phi}_a - \vec{\phi}_b$ and will likewise use $a^\mu_{ab} = a^\mu_a - a^\mu_b$.
Consequently fluctuations about this configuration can be thought of as being light with respect to this scale, or heavy and of this scale. Since we have a moduli space of vacua, there are light modes which describe the dynamics of these moduli, namely the long wavelength fluctuations in the parameters of the vacuum, $a^\mu_a$ and $\phi^I_a$, the diagonal components of the bosonic fields. Conversely the off-diagonal excitations of the fields are massive, with a scale set by $\sim | \vec{\phi}_{ab} |$. In the limit where this mass scale is suitably large, far out on the Coulomb branch, we might hope to integrate out the off diagonal degrees of freedom and generate an effective theory for the moduli fields. 

Now consider finite temperature $T$. Provided we consider $| \vec{\phi}_{ab}| \gg T$, then the same picture applies. We may separate the light degrees of freedom, which are still the fluctuations in $a^\mu_a$ and $\phi^I_a$, and consider their effective theory from integrating out the remaining off diagonal degrees of freedom. 
The difference is that now the moduli theory will itself be at finite temperature, and will receive additional temperature dependent corrections.

If the corrections to the classical moduli action from integration over the massive degrees of freedom 
are weak then the moduli would have a free thermal behaviour\footnote{For $p=0$ the free thermal energy diverges due of the large fluctuations of the zero mode similar to the perturbative limit (footnote \ref{fn-weak}).
Unlike the perturbative limit we do not know whether weak interactions on the Coulomb branch might regularise the thermal behaviour, and whether this state would be stable.
Hence the free thermal energy (\ref{eq:free-moduli}) may not be applicable for $p=0$. 
} 
\begin{eqnarray}
\epsilon \sim N T^{p+1} \; .
\label{eq:free-moduli}
\end{eqnarray}
However this free thermal behaviour, having entropy going as $O(N)$, would be dominated by the typical $O(N^2)$ entropy of large-$N$ gauge theories.
This may imply that this free thermal behaviour is unstable if we turn on the interactions for the moduli.

The loop expansion for the calculation of the interactions is controlled by the dimensionless couplings $\lambda_p/|\vec{\phi}_{ab}|^{3-p}$ and we emphasise that it can be computed for all $p$ (including the subtle $p \ge 3$ cases). It is important to understand that this computation, while weakly coupled, is quite distinct from the usual perturbation theory. For example, for $p \le 3$ here we may consistently have $\lambda_{eff} \gg 1$, and regain control provided that $\lambda_p/|\vec{\phi}_{ab}|^{3-p} \ll 1$, so that we are far enough out on the Coulomb branch. 
Integrating out the massive degrees of freedom, one obtains the finite temperature effective moduli theory,
\begin{align}
\label{Dp-moduli-action}
&S_{\text{eff}}^{\text{SYM}} = S_{\text{cl}}^{\text{SYM}}+S_{\text{one-loop}}^{\text{SYM}}+S_{\text{two-loop}}^{\text{SYM}}+ \cdots ,
\\
&S_{\text{one-loop}}^{\text{SYM}} =S_{\text{one-loop}}^{\text{non-thermal}}+ S_{\text{one-loop}}^{\text{thermal}}\,. \nonumber
\end{align}
Here $S_{\text{cl}}^{\text{SYM}}$ is the classical term directly from the original action (\ref{eq:SYM}),
\begin{align}
S_{\text{cl}}^{\text{SYM}} & = \frac{N}{\lambda_p} \int d \tau d^{p}x  \sum_{a=1}^N \left( \frac{1}{2} (\partial \vec{\phi}_a)^2 + \frac{1}{4} (F^{\mu\nu}_a)^2 + \text{fermions} \right) \nonumber \\
& =  \frac{N}{\lambda_p} \int d \tau d^{p}x  \sum_{a=1}^N \left(  \frac{1}{4} (F^{MN}_a)^2 + \text{fermions} \right) \label{classicalmoduli}
\end{align}
where $F_{\mu\nu}^a \equiv \partial_\mu a^a_{\nu}-\partial_{\nu} a_\mu^a$. 
Here we have introduced 10-dimensional SYM notation, where $F^{a}_{MN}$ ($M=(\mu,I)$) is the field strength of the $10$-dimensional SYM theory. Through classical dimensional reduction this can be viewed as the parent of our $(p+1)$-dimensional theory, where scalar fields are encoded in the extra gauge components in $10$-dimensions, $A^I=\phi^I$, so,
\begin{align}
F^{a}_{MN}F^{a}_{MN}=&  2  \partial_\mu \vec{\phi}_{a}\cdot \partial^\mu \vec{\phi}_{a} +  F^{a}_{\mu\nu}F^{a}_{\mu\nu}.
\end{align}

Since we are interested in finite temperature we consider the theory in Euclidean time, $\tau$, periodically identifying $\tau \sim \tau + 1/T$. Whilst there is no explicit temperature dependence in the classical part of the moduli action \eqref{classicalmoduli}, we emphasise that these light moduli fields `feel' the finite temperature due to this periodicity of Euclidean time. \footnote{
\label{ftnt-symmetric}
There is a subtlety that we have brushed over. The scalar part of the classical moduli action above appears to be a sigma model in the target space $\prod_N \mathbb{R}^{9-p}$. However, we have ignored the SYM gauge transformations that simply permute the diagonal components of the SYM matrix fields. Hence really we should identify the classical moduli theory under such permutations $(\vec{\phi}_a , a^\mu_a) \to (\vec{\phi}_{P_a} , a^\mu_{P_a})$, for $P$ a permutation $(1,2,\ldots,N) \to (P_1, P_2, \ldots, P_N)$. Doing so for all permutations, then the sigma model target space for the scalars becomes the symmetric product $S_N \mathbb{R}^{9-p}$. 
This will be relevant for $p=1$ \cite{Harvey:1995tg,Bershadsky:1995vm,Dijkgraaf:1997vv} in the later Section \ref{sec:D1}.
}

The weakly coupled massive modes yield loop corrections. $S_{\text{one-loop}}^{\text{SYM}}$ are the quantum and thermal corrections induced by the one-loop integral of the off-diagonal components.
The quantum contribution that arises at zero temperature, which we therefore term `non-thermal', takes the form (again using the 10-d SYM notation),
\footnote{This one-loop result is derived in Appendix B.}

\begin{align}
\label{eq:non-thermal}
S_{\text{one-loop}}^{\text{non-thermal}} =&  -\frac{  (2\pi)^{4-p} }{32 (7-p)\Omega_{8-p} } \nonumber \\
&\times  \int d\tau d^px \sum_{a<b}^{}  \frac{ 4 F^{ab}_{MN}F^{ab}_{NL}F^{ab}_{LK}F^{ab}_{KM} -  \left( F^{ab}_{MN} F^{ab}_{MN} \right)^2 }{ \left| \vec{\phi}_{ab} \right|^{7-p} } \left(1+ O\left( \frac{(F_{MN}^{ab})^2}{\phi_{ab}^4} \right) \right) 
\end{align}
for small ${(F^{ab}_{MN})^2}/{\phi_{ab}^4}$, and where again we use the notation $ F^{ab}_{MN} \equiv F^{a}_{MN}- F^{b}_{MN} $. 
Here
$\Omega_{8-p}(\equiv 2 \pi^{\frac{9-p}2}/\Gamma(\frac{9-p}2))$ is the volume of a unit $(8-p)$ sphere. We may explicitly expand in terms of ($p+1$)-fields,
\begin{align}
F^{ab}_{MN}F^{ab}_{NL}F^{ab}_{LK}F^{ab}_{KM}=&
2\left( \partial_\mu \vec{\phi}_{ab} \cdot \partial_\nu \vec{\phi}_{ab} \right)^2+4 \partial_\mu \vec{\phi}_{ab} \cdot \partial_\nu \vec{\phi}_{ab}F^{ab}_{\mu\rho}F^{ab}_{\nu\rho}
+ F^{ab}_{\mu\nu}F^{ab}_{\nu\rho}F^{ab}_{\rho\kappa}F^{ab}_{\kappa \mu} \,.
\end{align}
We note there is no explicit temperature dependence, but as with the classical terms periodicity of Euclidean time implicitly records the temperature.
An important comment is that these zero temperature corrections start at four derivative order, since supersymmetry protects the moduli from developing a potential, and maximal supersymmetry also protects the two derivative kinetic terms from corrections.

At one-loop the effective theory also receives corrections due to finite temperature. Since finite temperature manifestly breaks supersymmetry, the form of these corrections is more complicated, and there are corrections at zero, two, four derivative order and beyond, as computed in \cite{Wiseman:2013cda}. For example, the potential is the simplest correction at one-loop, and takes the form,
\begin{eqnarray}
\label{eq:thermal}
S_{\text{one-loop}}^{\text{thermal, pot'l}} & = &  - \frac{16}{ (2 \pi)^{p/2} } \int d\tau d^px \sum_{a<b}   \frac{ U_a U^\star_b + U_b U^\star_a }{ \beta^{1 + p} }  e^{- \beta |  \vec{\phi}_{ab} |} \left( \beta |  \vec{\phi}_{ab} | \right)^{p/2} 
\end{eqnarray}
where $U_a$ is the Polyakov loop around the Euclidean time circle, so that $U_a = e^{i \oint d a_a} = e^{i \oint d \tau a^0_a}$.\footnote{Note that for $p=0$ this potential term was computed in \cite{Ambjorn:1998zt}.} We see explicitly for this potential that it is proportional to $\exp(-\beta | \vec{\phi}_a-\vec{\phi}_b| )$ which is the Boltzmann factor of the off-diagonal components which have mass $\propto |\vec{\phi}_a-\vec{\phi}_b|$. Likewise, the contributions at higher derivative order are similarly controlled by the same Boltzmann factor dependence \cite{Wiseman:2013cda}.

As discussed above, at zero temperature this moduli action encodes the dynamics of $N$ nearly parallel D$p$-branes in flat 10-dimensional space, interacting weakly via the gravitational sector \cite{Morita:2013wfa}. In particular, if we consider zero temperature, and identify $X^I_a = 2 \pi \alpha' \phi^I_a$ as the transverse positions of the nearly coincident $N$ D$p$-branes, then the leading two derivative term above gives the dynamics due to the brane tensions, and the four derivative term accounts for classical exchange of the 10-d supergravity graviton, dilaton and form field. 
This agreement is of course well known \cite{Becker:1997xw}, resulting from non-renormalisation theorems, and is the basis for postulating the holographic correspondence between the gauge theory and a dual closed string theory.

At finite temperature we argued in \cite{Morita:2013wfa} that as usual for an interacting thermal system it should obey a virial relation, where the dynamics due to the leading two derivative kinetic term is balanced by the interaction terms. This assumes a good thermal vacuum exists and does not apply to the case where it is simply dominated by the leading free two derivative terms. The virial relation immediately implies that the theory will be strongly coupled, since one cannot assume the higher order terms are small, and in our moduli theory we have an infinite set of higher loop terms which we believe all become equally relevant.

The crucial point is that whilst a naive extrapolation of usual perturbation theory to strong coupling yields unphysical results, a naive extrapolation of the moduli theory to strong coupling reveals many of the features of black D$p$-brane dynamics.

\subsection{Recovering the dual black D$p$-brane phase \label{recoveringDp}}
\label{sec:Dp-brane-gravity}

The dual black D$p$-brane phase in IIA or IIB supergravity is the canonical strongly coupled phase of SYM, and in particular this phase applies in the 't Hooft regime. 
The black $p$-brane describes the thermal theory in the large $N$ limit, for effective coupling $\lambda_{eff} \sim O(N^0)$. However an analysis of $\alpha'$ corrections \cite{Itzhaki:1998dd} in supergravity shows that this gravitational description only applies for,
\begin{eqnarray}
\label{eq:blackbranevalidity}
1 \ll \lambda_{eff}^{\frac{1}{5-p}},
\end{eqnarray}
and as mentioned above for the opposite regime $\lambda_{eff}^{\frac{1}{5-p}} \ll 1$ the SYM becomes perturbative if $p<5$.

As we will discuss later, this phase in fact extends beyond this 't Hooft thermal regime even though for all $p$ the dual IIA or IIB black brane description eventually breaks down for $N^{\frac{2
}{7-p}} \ll \lambda_{eff}^{\frac{1}{5-p}}$ due to strong string coupling. For some $p$ there then is a phase transition to a new behaviour at sufficiently strong coupling.
To begin our discussion, we now review how the moduli theory may be used to see the existence of the dual black D$p$-brane phase directly in the SYM for $\lambda_{eff} \sim O(N^0)$.

In order to proceed we will assume that only a subset of the terms in the moduli action are required to describe the thermal state. We call these the `leading' contributions \cite{Morita:2013wfa}. By this we mean the lowest derivative terms at each loop order of the non-thermal corrections. We will assume that all higher derivative terms, 
controlled by $(F_{MN}^{ab})^2/\phi_{ab}^4$ 
at a given loop order, are negligible comparable to the lowest derivative term. Further we assume that the thermal corrections are subdominant to these leading non-thermal ones. We then argue that these assumptions are self-consistent within our extrapolation to strong coupling.

Next we assume that the gross properties of the thermal state scale are controlled by one physical scale, $\chi$, which we can think of as characterising the thermal vev of the scalars. In the spirit of a mean field approach, we then estimate the vev of a local operator by using the following rules.
Where one sees scalars in the vev, or differences between scalars, we estimate by the physical scale $\chi$,
\begin{eqnarray}
\vec{\phi}_a \sim \vec{\phi}_a - \vec{\phi}_b \sim \chi \,.
\end{eqnarray}
Derivatives are estimated using the thermal scale,
\begin{eqnarray}
\partial_\mu \sim \pi \, T
\end{eqnarray}
provided that only the Euclidean time direction is compact with anti-periodic fermion boundary conditions.
Thus we estimate $\partial_\mu \vec{\phi}_a \sim \pi \, T \, \chi$. 

We note that for $p \ne 3$ there is another dimensional scale in the problem as well as $T$, namely the coupling $\lambda_p$. The reason that we estimate time derivatives with the scale $T$, rather than with for example a scale derived from the coupling is that the thermodynamics is believed to occur in a scaling regime, where the temperature $T$ is far separated from the scale set by $\lambda_p$. Recall the 
regime of interest
is $1 \ll \lambda_{eff}^{\frac{1}{5-p}}$ with  $\lambda_{eff} = \lambda_p \, T^{p-3}$. 
Hence assuming the theory is not trivial (so there is no mass gap) at such temperatures, then the only relevant energy scale should be the temperature itself.

We make the same estimate for the gauge field, so that $\partial_\mu {a_\nu}_a \sim {F_{\mu\nu}}_a \sim \pi \, T \, \chi$.
In particular this means that where we see $F^{MN}_a$ or $F^{MN}_{ab}$ in a vev we make a replacement, 
\begin{eqnarray}
\label{eq:derivest}
F^{MN}_a \sim F^{MN}_{ab} \sim \pi \, T \, \chi \,.
\end{eqnarray}
We might be tempted to treat the gauge field without derivative in the same manner as the scalars, so that,
\begin{eqnarray}
\label{eq:aest}
{a}^\mu_a \sim \vec{a}^\mu_a - \vec{a}^\mu_b \sim \chi \,.
\end{eqnarray}
We note however, that the constant mode of the gauge field about the thermal circle has a potential from 
a different source in equation \eqref{eq:thermal} than the scalars, which obtain a potential directly from the non-thermal one-loop term \eqref{eq:non-thermal}. We will later argue that the potential  \eqref{eq:thermal} will be negligible in the regime we are interested in, and therefore the estimate above in \eqref{eq:aest} may not hold due to this lack of potential for the constant mode. Of course this does not affect the estimate for the derivatives of $a^\mu_a$ in \eqref{eq:derivest} which do not care about the constant mode.

A very important point to emphasise is that one may see above we are including factors of $\pi$ in our estimation, and we will later also consider geometric factors from sphere volumes. This is for the simple reason that it apparently reproduces the correct answer. At present we have little understanding for why, but presumably it is due to non-renormalisation properties of the theory. We emphasise that in estimating the derivative above it is natural to include a factor of $\pi$, since a Matsubara mode takes the form,
\begin{eqnarray}
\psi_n(\tau) = e^{2 \pi n i T \tau} \; , \quad \partial_\tau \psi_n = 2 \pi T n i \psi_n
\end{eqnarray}
and hence keeping factors of $\pi$ it is $(\pi T)$ that is 
associated to $\tau$ derivatives.

As we will see later, we will have factors of powers of $\pi$ in our expressions as well as $(8-p)$-sphere volumes, $\Omega_{8-p}$, originating from the non-thermal one-loop term in \eqref{eq:non-thermal}. Now again recalling $\Omega_{8-p}(\equiv 2 \pi^{\frac{9-p}2}/\Gamma(\frac{9-p}2))$, then we see for integer $p$ that $\Omega_{8-p}$ is simply a power of $\pi$ multiplied by a rational number, and therefore for integer $p$ there is no need to separately account for factors of $\pi$ and factors of $\Omega_{8-p}$, they may simply be combined into powers of $\pi$. We emphasise that in this work we will implicitly consider $p$ to be a continuous parameter that we may continue away from the integer values, and in this case the sphere volume involves a non-trivial factor due to the Gamma function, so that it is not simply powers of $\pi$ up to a rational number. Considering $p$ as a continuous parameter then it makes sense to treat factors of powers of $\pi$ and $\Omega_{8-p}$ separately as we shall do.

Lastly any sums over colour index are estimated in the obvious manner, $\sum_a \sim N$, $\sum_{a<b} \sim N^2$ etc..., where we focus on the large $N$ limit only.

Let us illustrate this by considering the operators of the classical and leading one-loop densities in the moduli action (ignoring fermions);
\begin{align}
\label{eq:leadingterms}
L_{0} & = \frac{N}{\lambda_p}  \sum_{a=1}^N  \frac{1}{4} (F^{MN}_a)^2 , \nonumber \\
L_{2} & = \frac{ (2\pi)^{4-p} }{32 (7-p) \Omega_{8-p} } \sum_{a<b}^{}  \frac{ 4 F^{ab}_{MN}F^{ab}_{NL}F^{ab}_{LK}F^{ab}_{KM} -  \left( F^{ab}_{MN} F^{ab}_{MN} \right)^2 }{ \left| \vec{\phi}_{ab} \right|^{7-p} } \,.
\end{align}
Firstly consider the (renormalised) finite temperature vev of $L_0$. We may use our rules to estimate,
\begin{align}
\left< L_{0}^{} \right> & = \left< \frac{N}{\lambda_p} \sum_{a=1}^N  \frac{1}{4} (F^{MN}_a)^2  \right> \nonumber \\
& \sim  \left< \frac{N}{\lambda_p} \sum_{a=1}^N (F^{MN}_a)^2  \right> \nonumber \\
& \sim \frac{N}{\lambda_p} \times N \times  ( {\pi}{T} \chi )^2 \; = \; \frac{N^2 \pi^2 T^2 \chi^2}{\lambda_p }
\end{align}
where we emphasise that we have dropped any rational constants (going from the first to the second lines), but have kept the transcendental geometric factors involving the $\pi$'s.
For $L_2$ we estimate the (renormalised) vev,
\begin{align}
\left< L_{2} \right>  & = \left<\frac{  (2\pi)^{4-p} }{32 (7-p)\Omega_{8-p} } \sum_{a<b}^{}  \frac{ 4 F^{ab}_{MN}F^{ab}_{NL}F^{ab}_{LK}F^{ab}_{KM} -  \left( F^{ab}_{MN} F^{ab}_{MN} \right)^2 }{ \left| \vec{\phi}_{ab} \right|^{7-p} }  \right> \nonumber \\
& \sim \qquad \frac{ \pi^{4-p}}{  \Omega_{8-p} } \quad \times \; N^2 \quad \times \quad \qquad \frac{ ( {\pi}{T} \chi )^4 }{\chi^{7-p} }  \qquad = \qquad \frac{ N^2 \pi^{8-p} T^4 }{ \Omega_{8-p}  \chi^{3-p}} \,.
\end{align}

Using these simple rules we argue that one can estimate the (renormalised) vevs of local operators. However, we have an undetermined scale $\chi$. This we fix by the assumption that the thermal state is strongly coupled and virialised. If we ignored all higher loop terms, and focus on the zero- and one-loop leading terms in the moduli action, we would obtain a (Euclidean) virial theorem,\footnote{This can be derived from the Lorentzian moduli theory in the usual manner and continued to Euclidean time, or obtained directly from the Euclidean path integral as a suitably renormalised Schwinger-Dyson equation associated to the scaling of the fields, $\vec{\phi}_a \to (1 + \epsilon ) \vec{\phi}_a$, $a^\mu_a \to (1 + \epsilon ) a^\mu_a$ for infinitesimal $\epsilon$.}
\begin{align}
\label{eq:virial}
\left<  \int d\tau d^px \left( 2 L_0 - (3-p) L_2 \right) \right> & = 0 \,.
\end{align}
Assuming spatial homogeneity this implies the finite temperature relation,
\begin{align}
\left<  L_0  \right>  \sim \left< L_2  \right> 
\end{align}
where we have dropped dependence on rational constants. Using our estimation methods for these vevs from above then implies,
 \begin{align}
 \label{eq:est1}
 \frac{N^2 \pi^2 T^2\chi^2}{\lambda_p } \sim  \frac{ N^2 \pi^{8-p} T^4}{ \Omega_{8-p}  \chi^{3-p}} \quad \implies \quad \chi^{5-p} \sim \frac{ \lambda_p \pi^{6-p} T^2}{ \Omega_{8-p}  } 
\end{align}
and hence fixes our unknown physical scale.

Thus at strong coupling we estimate that each leading term in the
classical and one-loop
 moduli action contributes with the same parametric form, 
 with the scalar and gauge fields contributing in the same way. 
 We have only considered the leading one-loop term, but as we have argued in \cite{Morita:2013wfa} we expect that all the leading higher loop terms in the moduli action also have precisely the same parametric dependence. 
\footnote{
Note that while the virial relation \eqref{eq:virial} becomes trivial for $p=3$ really we are concerned with strong coupling where all leading terms at arbitrary loop order contribute with the same parametric form, not just the first two terms.
}

The virialised state exhibits a low energy scale invariance at strong coupling, governed by the leading terms at all loop orders. As noted in \cite{Morita:2013wfa}, the classical and one-loop leading terms scale covariantly under the rigid transformation,
\begin{eqnarray}
\label{eq:scale}
(\tau, x^i) \to ( \Lambda^{-\frac{5-p}{3-p}} \tau, \Lambda^{-\frac{5-p}{3-p}} x^i ) \; , \quad \vec{\phi}_a \to \Lambda^\frac{2}{3-p} \vec{\phi}_a \; , \quad a^\mu_a \to \Lambda^\frac{2}{3-p} a^\mu_a
\end{eqnarray}
with,
\begin{eqnarray}
\label{eq:actscale}
L_0 \to \Lambda^\frac{2 (7-p)}{3-p} L_0 \; , \quad L_2 \to \Lambda^\frac{2 (7-p)}{3-p} L_2 \,.
\end{eqnarray}
The non-thermal corrections to the leading terms which we ignore, 
\begin{eqnarray}
\frac{ (\partial \phi)^2 }{\phi^4} \to \Lambda^2\frac{ (\partial \phi)^2 }{\phi^4} 
\end{eqnarray}
scale to zero as we take $\Lambda \to 0$. 
Note that the temperature scales as  $T \to \Lambda^{\frac{5-p}{3-p}} T$. Then the Boltzmann factor of the thermal corrections also scales to zero,
\begin{eqnarray}
e^{- \beta \phi } \to e^{- \beta \phi  \Lambda^{-1}}
\end{eqnarray}
in the $\Lambda \to 0$ limit. 
Based on this scaling property we can say that the strongly coupled regime is dominated by the terms which scale covariantly in the same manner as the classical and one-loop terms, and we expect that at each loop order there are such terms present in the non-thermal theory. 

Consider now the energy density of the theory. The (Euclidean) stress tensor of the Yang-Mills theory is,
\begin{align}
T_{\mu\nu} = \frac{N}{\lambda_p} \Tr{ \left[  \mathbf{F}_{\mu\alpha} \mathbf{F}_\nu^{~~\alpha} + D_\mu \mathbf{\Phi}^I D_\nu \mathbf{\Phi}^I - \eta_{\mu\nu} \left( \frac{1}{4} \mathbf {F}^2 + \frac{1}{2} (D \mathbf{\Phi}^I)^2 - \frac{1}{4}\left[ \mathbf{\Phi}^I, \mathbf{\Phi}^J \right]^2 \right)  \right] } + \mathrm{fermions} .
\end{align}
Just as with the Euclidean action where we could integrate out the off-diagonal modes to produce an effective action governing the moduli, we can consider the vev of the stress tensor in terms of these moduli degrees of freedom. At leading classical order since the moduli are diagonal all commutators vanish, and the classical contribution to the stress tensor is,
\begin{align}
\left< T_{\mu\nu} \right> = \left< \frac{N}{\lambda_p} \sum_a \left[  (F_a)_{\mu M}  (F_a)_\nu^{~~M} -  \frac{1}{4}  \eta_{\mu\nu}(F^{MN}_a)^2  \right] + \ldots \right>
\end{align}
where the ellipsis represents corrections from integration over the off-diagonal modes. We could equally obtain this expression by considering the stress tensor of the effective moduli action. In particular the terms given above are from the classical term, $\partial L_0 / \partial g_{\mu\nu}$, and the ellipsis above would be due to the higher loop terms, such as $\partial L_2 / \partial g_{\mu\nu}$.
The important point is that using our naive strong coupling extrapolation we can estimate this vev, to obtain,
\begin{align}
\label{eq:stresstensorest}
\left< T_{\mu\nu} \right> & \sim  \left< \frac{N}{\lambda_p} \sum_a \left[  (F_a)_{MN}  (F_a)_{PQ}  \right] + \ldots \right> \nonumber \\
& \sim \frac{N}{\lambda_p} \times N \times \left( {\pi}{T} \chi \right)^2 = \frac{N^2 \pi^2 T^2 \chi^2}{\lambda_p } 
\end{align}
which of course is the same estimate as for $\left< L_0 \right>$ and $\left< L_2 \right>$, essentially the statement that the vev of the 
Euclidean Lagrangian density is related to the free energy density,
which in this phase behaves in the same way as the energy density, and hence has the same estimate.

Then, using the fact that we have determined the scale $\chi$ using our strongly coupled virial assumption, \begin{align}
\label{eq:energydensity}
\left< T_{\mu\nu} \right> & \sim \frac{N^2 \pi^2 T^2}{\lambda_p } \left( \frac{ \lambda_p \pi^{6-p} T^2}{ \Omega_{8-p} }  \right)^{\frac{2}{5-p}} = N^2 \Omega_{8-p}^{-\frac{2}{5-p}} \pi^\frac{22- 4 p}{5 - p} \left( T \lambda_p^{- \frac{1}{3-p}} \right)^{\frac{ 2 (7-p) }{5 - p}} \lambda_p^\frac{1+p}{3-p}
\end{align}
which gives an estimate for both the energy and pressure of the strongly coupled thermal phase (we note this still correctly applies in the conformal case $p=3$).
\footnote{
Note that we have not considered or used any additional information from Lorentz structure, or assumptions of generalised conformal symmetry \cite{Jevicki:1998ub,Kanitscheider:2008kd} that would further constrain the vev of the stress tensor. 
Obviously here, the Lorentz symmetry, which is broken by finite temperature, would constrain the stress tensor to be static and have an isotropic pressure. The generalised conformal symmetry would constrain the trace of the stress tensor.
} 

Now we compare this stress tensor vev to that predicted by the thermodynamics of a black D$p$-brane in the gravity dual.
The string frame metric and dilaton for the black D$p$-brane is \cite{Itzhaki:1998dd},
\begin{align}
\label{eq:stringframe}
\alpha'^{-1} ds^2 & = \left( \frac{U^{7-p}}{ d_p \lambda_p } \right)^{\frac{1}{2}} \left( - f dt^2 + dx^i dx^i \right) +  \left( \frac{U^{7-p}}{ d_p \lambda_p } \right)^{-\frac{1}{2}} \left( \frac{1}{f} dU^2 + U^2 d \Omega_{8-p}^2 \right) \nonumber \\
e^\phi & = \left( 2 \pi \right)^{2-p} \frac{\lambda_p}{N} \left(  \frac{ d_p \lambda_p }{ U^{7-p} } \right)^{\frac{3-p}{4}} \nonumber \\
f & = 1 - \left( \frac{U_0}{U} \right)^{7-p} \; , \qquad d_p =  2^{7-2p} \pi^{\frac{9-3p}{2}} \Gamma\left( \frac{7-p}{2} \right)
\end{align}
where the radial coordinate $U$ has units of energy, and the horizon is located at $U = U_0$, and in addition the $p$-form field also has a non-trivial radial profile. Then using standard black hole 
thermodynamics one finds that the energy density $\epsilon$ in terms of entropy density $s$ is,
\begin{eqnarray}
\label{eq:entropy}
s = \frac{  \left( 2^{43-7 p}  \pi^{13-3p} (9-p)^{-(9-p)}  \Gamma\left( \frac{9-p}{2} \right)^2 \right)^\frac{1}{2(7-p)}  }{\sqrt{7-p}}  N^2 \left( \frac{\epsilon}{N^2 \lambda_p^\frac{1+p}{3-p}} \right)^\frac{9-p}{2(7-p)}  \lambda_p^\frac{p}{3-p}
\end{eqnarray}
which using the first law, and the relation,
\begin{eqnarray}
\Omega_{8-p} = \frac{ 2 \pi^{\frac{9-p}{2}} }{ \Gamma\left( \frac{9-p}{2} \right) } 
\end{eqnarray}
gives,
\begin{align}
\label{eq:gravDp}
\epsilon & = (9-p) \left( \frac{ 2^{31-5p}  \pi^{22-4p} }{ (7-p)^{3(7-p)} \Omega_{8-p}^2 }  \right)^\frac{1}{5-p}  N^2  (T \lambda_p^{-\frac{1}{3-p}})^{\frac{2( 7-p)}{5-p}} \lambda_p^\frac{1+p}{3-p} \nonumber \\
& = b_p \, N^2 \Omega_{8-p}^{-\frac{2}{5-p}} \pi^\frac{22- 4 p}{5 - p} \left( T \lambda_p^{- \frac{1}{3-p}} \right)^{\frac{ 2 (7-p) }{5 - p}} \lambda_p^\frac{1+p}{3-p} \;  , \quad b_p = (9-p) \left( \frac{ 2^{31-5p}  }{ (7-p)^{3(7-p)} }  \right)^\frac{1}{5-p} 
\end{align}
and we note that up to the factor $b_p$, which does not contain the geometric factors $\pi$ and $\Omega_{8-p}$, our estimate above from the stress tensor precisely reproduces this. We emphasise that this expression also applies in the case $p=3$.

We note that these Yang-Mills theories have been argued to enjoy a generalised conformal symmetry \cite{Jevicki:1998ub}. Such a symmetry, if indeed it survives in the quantum theory, is powerful in constraining non-local operators, such as correlation functions and transport coefficients \cite{Kanitscheider:2008kd}. However, as we see explicitly in \cite{Kanitscheider:2008kd}, it does not constrain the form of the entropy \eqref{eq:entropy} above beyond constraints already deduced from dimensional analysis. Hence it is important to note that our analysis is providing non-trivial information about the strong coupling behaviour that cannot simply be deduced from (generalised) symmetry of the full gauge theory.

\subsection{$\alpha'$ corrections to the D$p$-brane phase}

The transition from weak to strong coupling is at $\lambda_{eff} \sim O(1)$, and in the dual black D$p$-brane solution is understood to be due to $\alpha'$ corrections.  Considering the black brane \eqref{eq:stringframe} one finds,
\begin{align}
\label{eq:U0relation}
U_0^{5-p} = \frac{2^{13-2p} \pi^{11-2p} \lambda_p T^2}{(7-p)^3 \Omega_{8-p}}
\end{align}
and then we see that the radius, $R_{(8-p)}$, of the $(8-p)$-sphere at the horizon measured in string frame is,
\begin{align}
R_{(8-p)} = \sqrt{\alpha'} \left( \frac{d_p \lambda_p }{U_0^{3-p}} \right)^{\frac{1}{4}}
\end{align}
and hence for the $\alpha'$ corrections to be negligible we require $R_{(8-p)}$ to be large in string length units (which is sufficient for the string metric everywhere to be weakly curved in units of $\alpha'$), and hence,
\begin{align}
\label{eq:chi/T}
1 \ll \frac{ \lambda_p}{U_0^{3-p}} \qquad \implies \qquad 1 \ll  \lambda_p \left( \lambda_p T^2 \right)^{-\frac{3-p}{5-p}} = \lambda_{eff}^{\frac{2}{5-p}} .
\end{align}
Thus the condition for $\alpha'$ corrections to be negligible is,
\begin{align}
\label{eq:alpha}
1 \ll  \lambda_{eff}^{\frac{1}{5-p}}
\end{align}
and in the opposite regime where these corrections are strong one expects SYM perturbation theory to be valid if $p<5$ \cite{Itzhaki:1998dd}.
A challenge to our moduli description is that it should self consistently determine that the approximations made do not apply to this weakly coupled region of SYM, which as discussed in \ref{sec:weak} has a very different thermodynamic behaviour.

As emphasised in \cite{Wiseman:2013cda} for $p<3$, and then extended to the cases $p \ge 3$ in \cite{Morita:2013wla} the moduli theory indeed achieves this. 
Remember that our estimates above come from considering the leading terms in the moduli action. We have discarded corrections to the non-thermal action, controlled by $(\partial \phi_{ab})^2 / | \phi_{ab} |^4$, and also the thermal corrections which are controlled by the Boltzmann factor $e^{- | \phi_{ab} | / T}$. We may use our estimates to consider when this is self consistent, finding,
\begin{eqnarray}
\left< \frac{ (\partial \phi_{ab})^2 }{ | \phi_{ab} |^4 } \right>_T \sim \left(  \frac{ \pi T }{\chi} \right)^2  \; , \quad \left<  e^{- | \phi_{ab} | / T} \right>_T \sim  e^{ - \frac{ \chi }{T } } .
\end{eqnarray}
Hence both corrections are negligible when $1 \ll \chi/T$, and using our estimate for $\chi$, the corrections are negligible for,
\begin{eqnarray}
1 \ll \frac{\chi}{T} = \frac{1}{T} \left( \frac{ \lambda_p \pi^{6-p} T^2}{ \Omega_{8-p} }  \right)^{\frac{1}{5-p}} =  \left( \frac{ \pi^{6-p}}{ \Omega_{8-p} } \frac{ \lambda_p }{ T^{3-p} }  \right)^{\frac{1}{5-p}} 
\end{eqnarray}
and hence we recover precisely the condition that $1 \ll \lambda_{eff}^{{1}/{(5-p)}}$ in agreement with gravity for all $p$. Note we see that the dependence on $p$ in the power is quite non-trivially reproduced and is relevant when considering the subtle cases $p>3$ where the SYM itself is rather exotic.

Thus we assumed that only the leading terms were relevant in the strongly coupled regime, and then we see our estimates are only consistent with this if $1 \ll \lambda_{eff}^{1/(5-p)}$, although $O(N^0)$, which is precisely when the D$p$-brane phase is a good dual. For weaker coupling, $\alpha'$ corrections spoil this, and the gauge theory behaviour transitions over to the usual perturbative regime
\footnote{
For $p=5$, see Section \ref{sec-Hagedorn}.
At $p=6$, both gravity and gauge theory descriptions are good if $\lambda_{eff}\ll 1$. It may imply that these two states lie separately in the $\lambda_{eff}\ll 1$ regime \cite{Peet:1998wn}.}.
We note that numerical studies for D0 \cite{Catterall:2007fp, Hanada:2007ti} and D1-branes \cite{Catterall:2010fx} appear to be consistent with a smooth cross-over, and such a smooth transition has been argued in the context of the correspondence principle \cite{Horowitz:1996nw}.

Whilst the moduli theory predicts when $\alpha'$ corrections become important, an interesting question is whether it can quantitively describe these sub-leading corrections. Naively we would assume from \eqref{eq:non-thermal} that the fractional size of the corrections to the leading behaviour would go as $\sim (\partial \phi)^2/\phi^4 \sim (T/\chi)^2$. However in the cases $p=0$ and $p=3$ dual gravity arguments indicate that in fact the fractional size of the correction goes as the cubic power $\sim (T/\chi)^3$ \cite{Gubser:1998nz,Hanada:2008ez}.
Recall that our moduli theory is computed when it is weakly coupled, and then our estimates naively extrapolate its thermal behaviour to strong coupling. As we have said above, the fact that this apparently works for the leading behaviour must be due to non-renormalisation properties of the leading moduli theory. It seems unlikely to us that sub-leading terms  also enjoy such protection, and hence it is not surprising that we are unable to accurately estimate the size of the sub-leading corrections. That said, clearly this would be interesting to explore further.

\subsection{Predictions for other local operators}
\label{sec:predictions}

Beyond simply estimating the energy density in the thermal theory at strong coupling by naive extrapolation of our moduli theory applied to the stress tensor vev, we may also make predictions for the behaviour of various other local operators in the Yang-Mills theory. It would be very interesting to test these prediction where possible against calculations made in the 
string duals.

For example, we would estimate the vev of the operator,
\begin{align}
O_{F^2} = \left< \frac{1}{N} \Tr{ \left[  \mathbf{F}_{\mu\nu} \mathbf{F}^{\mu\nu} \right] } \right>  
\end{align}
by again considering the weakly coupled moduli approximation to it, and naively extrapolating to strong coupling, so,
\begin{align}
O_{F^2} & \sim \left< \frac{1}{N} \sum_a {F_a}_{\mu\nu} {F_a}^{\mu\nu} + \ldots \right> \sim \frac{1}{N} \times N \times ( \pi T \chi )^2 \nonumber \\
& \sim \pi^2 T^2 \left( \frac{ \lambda_p \pi^{6-p} T^2}{ \Omega_{8-p} }  \right)^{\frac{2}{5-p}} 
\end{align}
where the ellipsis represents higher loop corrections, which will all have the same parametric form after estimation, and hence we just use the first term as an estimate. Since we have argued that we estimate $\partial_\mu \phi^I$ and $F_{\mu\nu}$ in the same way, this implies,
\begin{align}
O_{(\partial \phi)^2} & = \left< \frac{1}{N} \Tr{ \left[ \partial^\mu \mathbf{\Phi}^I \partial_\mu \mathbf{\Phi}^I \right] } \right>  \nonumber \\
& \sim  \left< \frac{1}{N} \sum_a \partial^\mu \vec{\phi}_a \cdot  \partial_\mu \vec{\phi}_a  + \ldots \right> 
\sim  \pi^2 T^2 \left( \frac{ \lambda_p \pi^{6-p} T^2}{ \Omega_{8-p} }  \right)^{\frac{2}{5-p}} .
\end{align}
An important example is the scalar vev (with no derivatives),
\begin{align}
O_{(\Phi)^2} = \left< \frac{1}{N} \Tr{ \mathbf{\Phi}^I  \mathbf{\Phi}^I } \right>  
\end{align}
which is estimated at classical order as,
\begin{align}
O_{(\Phi)^2} & \sim \left< \frac{1}{N} \sum_a  \phi^I_a \phi^I_a + \ldots \right>  \sim  \frac{1}{N} \times N \chi^2  \nonumber \\
& \sim  \left( \frac{ \lambda_p \pi^{6-p} T^2}{ \Omega_{8-p} }  \right)^{\frac{2}{5-p}} 
\label{eq:Dp-scale}
\end{align}
and we will shortly argue that this provides a measure of the horizon radius in the dual gravity.

We note that large-$N$ factorisation of multi traces is obviously respected by these estimates. For example if we consider,
\begin{align}
O_{(\Phi)^2 (\Phi)^2} = \left< \frac{1}{N} \left( \Tr{ \mathbf{\Phi}^I  \mathbf{\Phi}^I } \right) \frac{1}{N}  \left( \Tr{ \mathbf{\Phi}^J  \mathbf{\Phi}^J } \right) \right>  
\end{align}
then the leading classical moduli contribution yields,
\begin{align}
O_{(\Phi)^2 (\Phi)^2} = \left<  \frac{1}{N} \left( \sum_a  \phi^I_a \phi^I_a \right)  \frac{1}{N} \left(\sum_b  \phi^J_b \phi^J_b \right)+ \ldots \right>  
\end{align}
and we will simply obtain an estimate where $O_{(\Phi)^2 (\Phi)^2} \sim (O_{(\Phi)^2})^2$ as we should expect at leading order at large $N$.

As final examples let us consider an operator quartic in the scalars, and one quartic in derivatives. Firstly,
\begin{align}
O_{(\partial \Phi)^4} = \left< \frac{1}{N} \Tr{ \left[  \partial_\mu \mathbf{\Phi}^I \partial_\nu \mathbf{\Phi}^I  \partial^\mu \mathbf{\Phi}^J \partial^\nu \mathbf{\Phi}^J \right] } \right> 
\end{align}
which to leading order in the moduli theory is,
\begin{align}
O_{(\partial \Phi)^4} & = \left< \frac{1}{N} \sum_a ( \partial_\mu \phi^I_a )  ( \partial_\nu \phi^I_a ) ( \partial_\mu \phi^J_a ) ( \partial_\nu \phi^J_a ) + \ldots \right>  \sim  \frac{1}{N} \times N ( \pi T \chi )^4  \nonumber \\
& \sim \pi^4 T^4 \left( \frac{ \lambda_p \pi^{6-p} T^2}{ \Omega_{8-p} }  \right)^{\frac{4}{5-p}} .
\end{align}
Secondly the higher derivative operator,
\begin{align}
O_{(\partial^2 \Phi)^2} = \left< \frac{1}{N} \Tr{ \left[  \partial_\mu \partial_\nu \mathbf{\Phi}^I \partial^\mu \partial^\nu \mathbf{\Phi}^I  \right] } \right> 
\end{align}
is estimated as,
\begin{align}
O_{(\partial^2 \Phi)^2} & = \left< \frac{1}{N} \sum_a (  \partial_\mu \partial_\nu \phi^I_a ) (  \partial^\mu \partial^\nu \phi^I_a ) + \ldots \right>  \sim  \frac{1}{N} \times N ( \pi^2 T^2 \chi )^2  \nonumber \\
& \sim \pi^4 T^4 \left( \frac{ \lambda_p \pi^{6-p} T^2}{ \Omega_{8-p} }  \right)^{\frac{2}{5-p}} .
\end{align}
Clearly one can consider many other examples. We argue that these predictions are highly non-trivial, giving precise parametric dependences of the vevs of various local operators, and it would be very interesting to find ways to test these by computing such quantities in the gravity duals if possible.

Finally, let us consider the Maldacena loop,
\begin{align}
W = \left< \frac{1}{N} \left| \Tr e^{ \oint  d\tau \left( i \mathbf{A}^\tau + n^I \mathbf{\Phi}^I \right) }  \right| \right>
\end{align}
where generally $n^I(\tau) \in \mathbb{R}^{9-p}$ is a unit vector valued over the time circle but we shall consider it is constant. 
Now at weak coupling the moduli theory gives,
\begin{align}
W = \left< \frac{1}{N} \left| \sum_a e^{\oint d\tau \left( i a^\tau_a + n^I \phi^I_a \right) + \ldots} \right| \right> .
\end{align}
The argument in the exponential comprises both an imaginary part from the gauge field, and a real part from the scalars. We have discussed above that an estimate for the gauge field constant mode is subtle. However for the real scalar part we estimate $\oint d\tau \left( n^I \phi^I_a  \right) \sim \frac{1}{T} \chi$. Without treating the gauge field constant mode, we can only use this real part to bound the estimate for $W$ as,
\begin{align}
W \lesssim \frac{1}{N} \times N \times e^{ \frac{\chi}{T} }
\end{align}
since the phase part may potentially contrive to cancel the contributions to $W$. Thus we conclude,
\begin{align}
\log W \lesssim \frac{\chi}{T} = \frac{1}{T} \left( \frac{ \lambda_p \pi^{6-p} T^2}{ \Omega_{8-p} }  \right)^{\frac{1}{5-p}} .
\end{align}
In fact this result may simply be checked from the dual black brane. The $p=0$ computation in  \cite{Hanada:2008gy} can be generalised straightforwardly to arbitrary $p$, where for the metric \eqref{eq:stringframe} one finds $\log{W} \sim U_0/(\pi T)$. From the relation \eqref{eq:U0relation} 
and the estimate \eqref{eq:est1} we see that $U_0 \sim \pi \chi$. Using this we see the gravity calculation of $\log{W}$ precisely agrees with our estimate above $\log{W} \sim \chi/T$, including the $\pi$ and sphere volume factors. 

\subsection{Horizon size}
\label{sec:horizonsize}

In order to discuss the horizon radius of the black brane dual we need a measure of it. The most natural measure is the radius of the horizon as measured by the branes themselves.
As discussed in \cite{Morita:2013wfa}, a probe brane that is nearly parallel to the $N$ branes giving rise to the metric \eqref{eq:stringframe} sees the transverse space,
\begin{align}
ds^2_{brane} & = \frac{1}{f} dz^2 + z^2 d \Omega_{8-p}^2 \; , \quad f = 1 - \left( \frac{z_h}{z} \right)^{7-p}
\end{align}
where $z = \alpha' U$ and $z_h = \alpha' U_0$. Note that this is the superspace metric for the moduli that describe the probe brane embedding into the background \eqref{eq:stringframe}, and is neither a projection of the Einstein frame, nor the string frame metric, but is related by additional powers of $e^\phi$. 
In this natural transverse brane metric the proper radius of the horizon is $z_h$.

As emphasised in \cite{Morita:2013wfa}, the relation between the transverse probe brane positions $X^I_a = 2 \pi \alpha' \phi^I_a$ at weak coupling and the Yang-Mills moduli $\phi^I_a$, is strongly suggestive that at strong coupling the size of the black brane horizon, $z_h$, as measured by the branes in their natural metric for the transverse space, is given in terms of the thermal vev of the scalars as,
\begin{eqnarray}
\label{eq:horizsize}
z_h^2 \sim  \pi^2 {\alpha'}^2 \left<  \frac{1}{N}  \Tr{ \mathbf{\Phi}^I \mathbf{\Phi}^I } \right> =  \pi^2 {\alpha'}^2 \left<  \frac{1}{N} \sum_a \vec{\phi}_a \cdot \vec{\phi}_a  + \ldots \right>
\end{eqnarray}
where $\sim$ indicates equality including dependence on $N$, dimensional parameters and $\pi$'s and sphere volumes, but not including possible non-transcendental constants.

Having already argued that our estimates should imply, 
\begin{align}
O_{(\Phi)^2} = \left< \frac{1}{N} \Tr{ \mathbf{\Phi}^I  \mathbf{\Phi}^I } \right>  \sim  \left( \frac{ \lambda_p \pi^{6-p} T^2}{ \Omega_{8-p} }  \right)^{\frac{2}{5-p}} 
\end{align}
then we conclude that,
\begin{align}
z_h^2 & \sim  \pi^2 \alpha'^2   \left( \frac{ \lambda_p \pi^{6-p} T^2}{ \Omega_{8-p} }  \right)^{\frac{2}{5-p}} .
\label{eq:phi}
\end{align}
Up to a remaining numerical factor (not involving $\pi$ or sphere volumes) this indeed gives the correct  relation for $z_h$ 
in the black brane solution, which we see directly 
in gravity
using $z_h = \alpha' U_0$ and the black brane relation \eqref{eq:U0relation}.

\subsection{The Hagedorn case $p=5$}
\label{sec-Hagedorn}

The case $p=5$ exhibits Hagedorn behaviour, where the temperature is constant, yet the energy density may vary. We have to apply our estimates more carefully in this special case, but we find they still correctly work.

Firstly our estimate in \eqref{eq:est1} for $p=5$ yields the Hagedorn temperature itself,
\begin{eqnarray}
T^2 \sim \frac{\pi}{\lambda_5} \,.
\end{eqnarray}
Thus the temperature is constant and the scale $\chi$ giving the vev of the scalars is now the variable in the thermodynamics. Then from \eqref{eq:stresstensorest}, we obtain,
\begin{eqnarray}
\left< T_{\mu\nu} \right> \sim \frac{1}{\lambda_5} N^2 \pi^2 T^2 \chi^2 =  \frac{1}{\lambda_{5}^2} N^2 \pi^3 \chi^2 
\end{eqnarray}
which is a function of $\chi$. Hence we relate the energy density and pressure to the vev of the scalars, determined by $\chi$.
Note that from the above, $z_h \sim \pi \alpha' \chi$, and using this for the stress tensor vev above correctly reproduces the black hole energy density and pressure as a function of horizon size $z_h$.

%---------------------------------------------------------------
\section{SYM on a spatial circle}
\label{sec:circle}
%---------------------------------------------------------------

We now consider the super Yang-Mills where we compactify a spatial direction. In this case dual gravity arguments indicate interesting new phase structure in the strongly coupled 't Hooft regime, where $1 \ll \lambda_{eff}^{1/(5-p)}$ but still $\sim O(N^0)$.

\subsection{D$p$-branes winding a circle}
\label{sec-Dp-winding-circle}

In the $N$ D$p$-brane system,
suppose that one of the spatial directions (say $x_p$) in the worldvolume is compactified to have radius $L$. If we took it to have anti-periodic boundary conditions for fermions, then in the Euclidean picture it would manifest analogously to the periodic Euclidean time direction, and hence would yield terms that are exponentially suppressed in  $L | \phi_{ab} | \gg 1$, of precisely the form of the thermal corrections above but with $\beta \to 2\pi L$. 
However, we will focus on the more interesting case, where the fermion boundary conditions are periodic. In this case, as we discuss in detail in Appendix \ref{app:GL}, in fact at one loop exponentially suppressed corrections in  $L | \phi_{ab} | \gg 1$ are generated, but the difference with the anti-periodic case is that instead of correcting all derivative orders,  they correct only four derivatives and higher. Indeed the form of the four derivative correction which we compute in the Appendix at one loop is,
\footnote{
This one-loop term  is governed by a harmonic potential
in a transverse space which includes an extra compact direction, with coordinate $a^p$, with radius inversely related to $L$. This is of course the harmonic interaction mediating gravity exchange between D$(p-1)$-branes derived from the D$p$-brane dynamics after a T-duality on the compact circle so that its radius $L \to \alpha'/L$.
}
\begin{align}
{S}^{E,4}_{compact}  = & - \int d\tau d^{p-1}x \frac{1}{2 \pi L} \int_0^{2\pi L}dx_p \sum_{a<b}  \sum_{n \in \mathbb{Z} }  \frac{ (2\pi)^{5-p}}{32 (8-p)  \Omega_{9-p} }
 \frac{ 4 F^{ab}_{MN}F^{ab}_{NL}F^{ab}_{LK}F^{ab}_{KM} -  \left( F^{ab}_{MN} F^{ab}_{MN} \right)^2 }{ \left( \left( \frac{ n}{L} - a_{ab}^p \right)^2 +  \left| \vec{\phi}_{ab} \right|^2  \right)^{\frac{8-p}{2}} } \,.
 \label{eq:int-compact}
\end{align}
There are then two interesting limits to this one-loop term.
Firstly when $L | \phi_{ab} | \gg 1$  this harmonic potential can be expanded to give our usual one in $(9-p)$ transverse dimensions (\ref{eq:non-thermal}), plus exponentially suppressed corrections in $L | \phi_{ab} |$;
\begin{eqnarray}
{S}^{E,4}_{compact}  & = & -  \int d\tau d^{p-1}x \int_0^{2\pi L}dx_p \sum_{a<b} \frac{  (2\pi)^{4-p}}{32 (7-p) \Omega_{8-p} }
 \frac{ 4 F^{ab}_{MN}F^{ab}_{NL}F^{ab}_{LK}F^{ab}_{KM} -  \left( F^{ab}_{MN} F^{ab}_{MN} \right)^2  }{  \left| \vec{\phi}_{ab} \right|^{ 7-p } }  \nonumber \\
 && \qquad \qquad \qquad \times \left( 1 + e^{- 2\pi L \left| \phi_{ab} \right| } \left(2\pi L \left| \phi_{ab} \right|  \right)^{\frac{6-p}{2}} \cos{ \left(2\pi L a^p_{ab} \right) } + \ldots \right) .
\end{eqnarray}
Now consider the thermodynamics of this system by using our usual estimates. A key assumption is that for periodic boundary conditions on the spatial circle we should still estimate $\partial \sim \pi T$. Obviously if $\beta = 1/T \ll L$ then temperature is the dominant energy scale. However, for $\beta \gg L$ one might imagine that the energy scale $1/L$ should be used to estimate the derivatives. Our use of $T$ is precisely related to the fact that we are assuming a non-trivial scaling dynamics at low temperatures, and indeed for temperatures well below the scale $\sim 1/L$, and therefore it should be this lowest scale that determines the behaviour of quantities such as derivatives.
\footnote{If we had taken anti-periodic boundary conditions on a spatial circle, then we would not obtain such a non-trivial dynamics at low temperature, and this would be reflected by
estimating 
$\partial \sim \pi \max\left( \frac{1}{\beta} , \frac{1}{2\pi L} \right)$, 
and a transition in behaviour
will occur at $2\pi L=\beta$ where the lowest scale is flipped.
A corresponding phase transition does indeed occur in the dual supergravity \cite{Horowitz:1998ha, Aharony:2005ew, Mandal:2011ws}.
}

We obtain our usual black D$p$-brane behaviour if we treat the correction term as negligible. However, we must check to ensure this assumption is self consistent, which requires that $L \left| \phi_{ab}  \right| \gg 1$ in the expression above in order to suppress the correction term. Our estimates yield,
\begin{eqnarray}
L \left| \phi_{ab} \right|  \sim L \chi \sim L ( \lambda_p T^2 )^{\frac{1}{5-p}} 
\end{eqnarray}
where since we just wish to determine the regime where the corrections are negligible we ignore geometric factors. Hence this implies that a deviation from the D$p$-brane behaviour (as detailed in Section \ref{recoveringDp}) will occur when,
\begin{eqnarray}
L \sim  ( \lambda_p T^2 )^{-\frac{1}{5-p}} .
\end{eqnarray}
This indeed agrees with the dual closed string predictions \cite{Susskind:1997dr, Barbon:1998cr, Aharony:2004ig, Aharony:2005ew, Mandal:2011hb}.  Defining the critical length for a given temperature as $L_c =  ( \lambda_p T^2 )^{-\frac{1}{5-p}}$, then precisely when $L \sim L_c$ string winding modes about the compact spatial direction become unstable. These can be understood by performing a T-duality on the circle, so that one is considering a black D$(p-1)$-brane \emph{smeared} over the circle. Since the thermodynamics of such a black brane is invariant under T-duality, this smeared D$(p-1)$-brane has the same behaviour as the original black D$p$-brane that wraps the circle. However, such a smeared solution becomes classically unstable in gravity to localising its horizon on the compact circle (the Gregory-Laflamme instability \cite{Gregory:1993vy, Horowitz:2011cq}), and this yields a
phase
transition to a new behaviour dual to a black D$(p-1)$-brane which is \emph{localised} on the circle. 
\footnote{
Evidence for this large $N$ phase transition has been found in direct numerical studies for $p=1$ \cite{Catterall:2010fx}.
Phase transitions in related Yang-Mills theories have been studied in \cite{Aharony:2004ig, Kawahara:2007fn, Mandal:2009vz, Azuma:2014cfa}. 
}
For $L \ll L_c$ the horizon of this black D$(p-1)$-brane is much smaller than the size of the T-dual circle (length $\alpha'/L$) and so the energy density of the gravity solution (and hence the SYM, recalling thermodynamics is T-duality invariant) behaves as that of a 
black D$(p-1)$-brane in a non-compact transverse space. 
More precisely, the energy density of the SYM \emph{integrated over the compact circle} should give the energy density of a non-compact black D$(p-1)$-brane solution.
So for $L \ll L_c$ we expect $\oint_0^{2 \pi L} dx_p \, \epsilon$ to be given by equation \eqref{eq:gravDp} with the replacements $p \to p-1$ where the resulting parameter $\lambda_{p-1}$ after this replacement is related by dimensional reduction to the original Yang-Mills coupling so $\lambda_{p} = 2 \pi L \lambda_{p-1}$.
\footnote{
We can see how this picture works more explicitly by recalling that in Section \ref{sec:horizonsize} we used our discussions in \cite{Morita:2013wfa} to argue the horizon size should be estimated by the scalar vev, with specific factors of $\alpha'$ and $\pi$ as in equation \eqref{eq:horizsize}. Then 
at the transition point $L \chi \sim 1$
we find that (dropping geometric factors),
\begin{eqnarray}
z_h \sim \alpha' \chi \sim \frac{\alpha'}{L} = L_{(T)}
\end{eqnarray}
and hence the horizon size is of order the radius of the T-dual circle, $L_{(T)} = \alpha' / L$, it is smeared over. This is precisely when a Gregory-Laflamme transition occurs.
}

One important comment is that this transition from D$p$-brane to D$(p-1)$-brane behaviour is highly non-trivial in the dual description of the SYM. Firstly it requires a T-dual frame to examine the gravity, and secondly it involves the Gregory-Laflamme transition. We emphasise that it does not simply arise from a `dimensional reduction' of a black D$p$-brane, which since it is wrapping the circle, cannot be reduced to give an appropriate D$(p-1)$-brane solution.

We have seen above that the correction to our moduli theory in  \eqref{eq:int-compact} in the limit $L \gg L_c$ precisely predicts the transition in behaviour. In fact as we shall now argue, we may also estimate the D$(p-1)$-brane behaviour for $L \ll L_c$ from our moduli action in the compact case \eqref{eq:int-compact}. Above the transition for $L \gg L_c$, we have argued $L \left| \phi_{ab}  \right| \gg 1$ and hence the corrections in
\eqref{eq:int-compact} due to the presence of the circle are negligible. Well below the transition for $L \ll L_c$  in the new phase with D$(p-1)$ behaviour, we expect,
\begin{eqnarray}
L \left| \phi^I_{ab} \right| \, , \;  L \left| a^p_{ab} \right| \ll 1
\end{eqnarray}
where we assume $\phi^I_{ab}$ and $a^p_{ab}$ have the same estimate. Note that previously in Section \ref{sec:Dp-brane-gravity} we have estimated the gradients $\partial \phi^I_a$ in the same way as the gradients of the gauge fields, ${F_{\mu\nu}}_a$, but were careful to point out that an estimate for the gauge field itself was subtle as there are no explicit potential terms for the gauge field in the leading action. However, we  see in the leading action with a compact circle now $a^p_a$ does have an explicit potential at 1-loop, and hence we are indeed justified in making the estimate $\phi^I_{ab} \sim a^p_{ab} \sim \chi$ for at least this $p$-component of the gauge field.

In the case $L \ll L_c$  
we see that only the $n=0$ term in the sum in \eqref{eq:int-compact} is important, and hence the classical and one-loop terms of the moduli action can be simplified to,
\begin{align}
S^2_{compact} =& \frac{1}{2 \pi L}  \int_0^{2\pi L}dx_p \frac{2 \pi L N}{\lambda_p} \int d \tau d^{p-1}x   \sum_{a=1}^N \left(  \frac{1}{4} (F^{MN}_a)^2  \right) \nonumber \\
{S}^{E,4}_{compact}  = & - \frac{1}{2 \pi L}  \int_0^{2\pi L}dx_p \int d\tau d^{p-1}x \sum_{a<b}   \frac{ (2\pi)^{5-p}}{32 (8-p)  \Omega_{9-p} } 
 \frac{ 4 F^{ab}_{MN}F^{ab}_{NL}F^{ab}_{LK}F^{ab}_{KM} -  \left( F^{ab}_{MN} F^{ab}_{MN} \right)^2 }{ \left( \left( a_{ab}^p \right)^2 +  \left| \vec{\phi}_{ab} \right|^2  \right)^{\frac{8-p}{2}}  } \,.
 \end{align}
We may trivially rewrite these expressions as,
 \begin{align}
S^2_{compact} =& \frac{1}{2 \pi L}  \int_0^{2\pi L}dx_p \frac{N}{\lambda_{p-1}} \int d \tau d^{p-1}x   \sum_{a=1}^N \left(  \frac{1}{4} (F^{MN}_a)^2  \right) \nonumber \\
{S}^{E,4}_{compact}  = & - \frac{1}{2 \pi L}  \int_0^{2\pi L}dx_p \int d\tau d^{p-1}x \sum_{a<b}   \frac{ (2\pi)^{4-(p-1)}}{32 (7-(p-1))  \Omega_{8-(p-1)} } 
 \frac{ 4 F^{ab}_{MN}F^{ab}_{NL}F^{ab}_{LK}F^{ab}_{KM} -  \left( F^{ab}_{MN} F^{ab}_{MN} \right)^2 }{  \left| \vec{\phi}'_{ab} \right|^{7-(p-1)}  }
 \label{eq:moduli-T-dual}
 \end{align}
 where $\lambda_{p} = 2 \pi L \lambda_{p-1}$, and $\vec{\phi'}_a = ( \vec{\phi}_a , a^p_a )$.
We immediately see that apart from the integral over $x^p$ in both terms, structurally the moduli action is that for SYM in one lower dimension, $p \to (p-1)$, where the gauge field component $a^p_a$ in that direction is now playing the role of the extra scalar, $\phi^{9-(p-1)}_a$. Hence our estimation now yields the correct answer. Explicitly, writing $\tilde{p} = p-1$, our virial relation,
\begin{eqnarray}
\left<  \frac{N}{\lambda_{\tilde{p}}}\sum_{a=1}^N \left(  \frac{1}{4} (F^{MN}_a)^2  \right) \right> \sim \left<  \sum_{a<b}   \frac{ (2\pi)^{4-\tilde{p}}}{32 (7-\tilde{p})  \Omega_{8-\tilde{p}} } 
 \frac{ 4 F^{ab}_{MN}F^{ab}_{NL}F^{ab}_{LK}F^{ab}_{KM} -  \left( F^{ab}_{MN} F^{ab}_{MN} \right)^2 }{  \left| \vec{\phi}'_{ab} \right|^{7-\tilde{p}}  } \right>
\end{eqnarray}
is estimated as,
\begin{eqnarray}
\frac{N^2}{\lambda_{\tilde{p}}} ( \pi T \chi )^2 \sim N^2  \frac{ \pi^{4-\tilde{p}}}{ \Omega_{8-\tilde{p}} } 
 \frac{ ( \pi T \chi )^4 }{  \chi^{7-\tilde{p}}  } \quad \implies  \quad  \chi^{5-\tilde{p}}  \sim \lambda_{\tilde{p}}  \frac{ \pi^{6-\tilde{p}} T^2}{ \Omega_{8-\tilde{p}} } 
\end{eqnarray}
and from earlier \eqref{eq:stresstensorest} the SYM stress tensor gives the vev of the energy density, $\epsilon$, to be,
\begin{eqnarray}
\epsilon \sim \frac{N^2 \pi^2 T^2 \chi^2}{\lambda_p } 
\sim \frac{1}{\pi L} \frac{N^2 \pi^2 T^2}{\lambda_{\tilde{p}} } \left( \lambda_{\tilde{p}}  \frac{ \pi^{6-\tilde{p}} T^2 }{ \Omega_{8-\tilde{p}} }  \right)^{\frac{2}{5 - \tilde{p}}} .
\end{eqnarray}
Then the energy density integrated over the circle 
(assuming it to be homogeneous)
will be,
\begin{eqnarray}
\int_0^{2 \pi L} dx_p \epsilon \sim 2 \pi L \epsilon  
\sim  \frac{N^2 \pi^2 T^2}{\lambda_{\tilde{p}} } \left( \lambda_{\tilde{p}}  \frac{ \pi^{6-\tilde{p}} T^2 }{ \Omega_{8-\tilde{p}} }  \right)^{\frac{2}{5 - \tilde{p}}} 
\end{eqnarray}
which recalling the earlier relation for the D$p$-brane energy density, \eqref{eq:energydensity}, shows that $\int_0^{2 \pi L} dx_p \epsilon$ does indeed reproduce the energy density of a D$\tilde{p}$-brane, where of course $\tilde{p} = p - 1$.

So we have seen that considering a spatial direction to be compact, for large $L \gg L_c$ introduces negligible corrections to the leading terms in the moduli action. The terms grow in our estimates, and when $L \sim L_c$ the terms are $\sim O(1)$ corrections, and thus the behaviour must change. This is precisely where in the T-dual gravity picture a Gregory-Laflamme transition occurs. Furthermore, we see that for small $L \ll L_c$, this T-dual picture predicts a D$(p-1)$-brane behaviour, and indeed our moduli theory exactly agrees with this.

Apart from the details of the thermal behaviour near $L \sim L_c$ our moduli estimates, using the full action computed for the compact circle \eqref{eq:int-compact}, are therefore able to reproduce both the thermodynamics of the small and large $L$ phases. We emphasise that just as in Section \ref{sec:predictions} we demonstrated how to estimate various local operator vevs, this can be extended simply to the case with compact circle. Obviously for $L \gg L_c$, we will just obtain the same estimates for these operators. However, for $L \ll L_c$, it should be clear that we will obtain different estimates, but they will simply be the same ones but with $p \to \tilde{p} = p-1$.

Finally we note that in \cite{Taylor:1996ik} it was argued that $(p+1)$-SYM compactified on one spatial circle (with periodic fermion boundary conditions) is equivalent to a $p$-dimensional SYM theory with non-compact spatial directions, but with a different gauge group, related to that of the $(p+1)$-d theory after an orbifolding. This gives the action of T-duality on the circle from the D$p$-brane world volume SYM perspective, giving the dual world volume theory of the D$(p-1)$-branes. We note that in this theory the moduli action can be computed and, by construction, gives precisely the same answer as \eqref{eq:int-compact}. From the point of view of the $p$-dimensional SYM with orbifold gauge group, there is the same phase transition, which is now interpreted simply as a Gregory-Laflamme transition in the compact direction of the transverse space of the D$(p-1)$-branes.

%---------------------------------------------------------------
\section{Very strong coupling: beyond the 't Hooft regime}
\label{sec:corrections}
%---------------------------------------------------------------

Up to this point we have applied our estimates in the 't Hooft regime where $\lambda_{eff} = \lambda_p / T^{3-p} \sim O(N^0)$ in the large $N$ limit. We have seen that the moduli theory, computed in its weakly coupled limit, correctly estimates the dual black brane behaviour when naively extrapolated to strong coupling.

Furthermore, it correctly encodes when $\alpha'$ corrections become important and hence the behaviour crosses over to thermal SYM at weak coupling, via corrections to the leading terms in the moduli action. 
Introducing a compact spatial circle radius $L$ we may again compute the moduli action when it is weakly coupled, and again the corrections to the leading moduli action \eqref{eq:leadingterms} correctly predict the transition at radius $L_c$ for fixed temperature in behaviour for both anti-periodic fermion boundary conditions (where $L_c \sim 1/T$), or the more interesting case of periodic conditions (where $L_c \sim ( \lambda_p T^2 )^{-\frac{1}{5-p}}$). 

In this section we wish to emphasise one of the most interesting parts of using naive strong coupling extrapolation of the moduli theory. For simplicity we now assume our spatial directions are non-compact. Then the strongly coupled phase dual to the D$p$-brane has the thermodynamic behaviour as in \eqref{eq:gravDp} in the 't Hooft regime. A natural question is what happens as we go to even stronger coupling, so $N^q \ll \lambda^{\frac{1}{5-p}}$ for $q>0$ - does this behaviour persist? 

The first thing to emphasise is that the dual black D$p$-brane description breaks down at strong coupling as the IIA/B string loop corrections become large at the black brane horizon. The condition that the dilaton becomes of order one occurs when,
\begin{eqnarray}
\lambda_{eff}^{\frac{1}{5-p}} \sim N^{\frac{2}{7-p}}
\end{eqnarray}
at a dimensionless temperature $\tilde{T} = T \lambda_p^{-1/(3-p)} = \lambda_{eff}^{-1/(3-p)}$ which we call $\tilde{T}_{Dp}$, so,
\begin{eqnarray}
\tilde{T}_{Dp} \sim N^{-\frac{2 (5-p)}{(7-p)(3-p)}} .
\end{eqnarray}
Beyond this coupling and temperature then this IIA or IIB gravity dual description breaks down. However, we note that this scale does not appear in the moduli theory, and hence we might imagine that the estimates for the thermal behaviour still persist even beyond this point. Indeed this is the case. We regard this as one of the most powerful aspects of the moduli approach as we will now describe.

In {all} the cases of $p = 0 , \ldots , 6$, when one 
pushes
the temperature past $\tilde{T}_{Dp}$ to stronger coupling, one is able to change to an alternate gravity dual. For $p$ even, this is found by uplifting the IIA black D$p$-brane to M-theory. Since such an oxidation of a black brane does not change the thermodynamics, we see that while the description requires a new gravity dual, in fact the thermodynamics are not changed. In the cases where $p$ is odd, one may perform a IIB S-duality, $g_s \to 1/g_s$, on the black D$p$-brane. The large string coupling near the horizon is now controlled in the new S-dual frame. The type of black brane changes, so D1 $\to$ F1 and D5 $\to$ NS5, with D3 being self dual. However, the transformation of the black brane solution under S-duality does not change the thermodynamics. 

Hence extrapolating the moduli to strong coupling and estimating the thermodynamics not only gives the correct behaviour of the dual black D$p$-brane, but also gives the correct behaviour going to stronger coupling where the black D$p$-brane description breaks down.

Consider the odd $p$ cases. The reason this works is of course that the moduli theory deduced from the SYM describes not only the D$p$-branes, but is S-duality invariant, and thus is also the moduli theory that describes the S-dual  black branes in the sense of \cite{Morita:2013wfa}. Hence the same estimates give the correct behaviour of this S-dual black brane.
Consider only the scalar degrees of freedom of the leading moduli theory, written in terms of the gravitational variables $X^I_a$ (given in terms of Yang-Mills variables earlier in Section \ref{sec:horizonsize}), $\mu=(2\pi)^{-p}g_s^{-1}{\alpha'}^{-\frac{1+p}{2}}$ and $2\kappa^2= (2\pi)^7 g_s^2 {\alpha'}^4$, which up to one loop are,
\begin{align}
S^{\text{scalar}} & = \int d\tau dx^p \frac{\mu}{2} \sum_{a=1}^N (\partial_\mu \vec{X}_a)^2
\nonumber \\
& \qquad - \frac{\mu^2 \kappa^2}{4 ( 7- p ) \Omega_{8-p}} \sum_{a<b} \frac{2 \left( \partial_\mu \vec{X}_{ab} \cdot \partial_\nu \vec{X}_{ab} \right)^2 - \left( \partial_\mu \vec{X}_{ab} \cdot \partial^\mu \vec{X}_{ab} \right)^2 }{ \left| \vec{X}_{ab} \right|^{7-p} } \nonumber \\
& \qquad + O(\kappa^4)  
\label{moduli-string-frame}
\end{align}
and precisely give the 
weakly coupled dynamics of nearly parallel $p$-branes
 in 10-dimensions as discussed in \cite{Morita:2013wfa}. Under an appropriate re-identification of $\mu$ and $\kappa^2$ in terms of string theory parameters the action also describes the S-dual brane. 
For example, in the case $p=1$ for the black D$1$-brane in IIB we have $\mu \sim 1/(g_s \alpha')$ and $\kappa^2 \sim g_s^2 \alpha'^4$. 
However following \cite{Morita:2013wfa} the S-dual brane, a black F1-string, has the same moduli action with the same $\mu$ and $\kappa^2$, now regarded as $\mu \sim 1/(\alpha'^{F1})$ and $\kappa^2 \sim (g_s^{F1})^2 (\alpha'^{F1})^4$ in terms of the S-dual IIB string parameters $g_s^{F1} =1/g_s $ and $\alpha'^{F1}=g_s \alpha'$.
A detailed discussion of the various cases of $p$ is given as part of Appendix \ref{app:fullstructure}.

We now focus on what happens at even stronger coupling. In some cases the M-theory descriptions, or S-dual IIB black brane descriptions predict a new phase transition, or a new physical behaviour. We will see that in certain cases this new physical behaviour can be understood precisely as being due to new sources of corrections to the moduli theory that become important. In other cases we currently have only a qualitative understanding for why the moduli description must change.

One interesting point to note is that the weakly coupled moduli theory itself has a thermal behaviour $\epsilon \sim N T^{p+1}$ \eqref{eq:free-moduli}, \footnote{As mentioned earlier, for $p=0$ there are subtleties associated to IR divergences.}
while the strong coupling estimates for the moduli theory give the behaviour of a D$p$-brane \eqref{eq:gravDp} $\epsilon \sim N^2 \tilde{T}^{2(7-p)/(5-p)} \lambda^{(1+p)/(3-p)}$.
These two behaviours coincide at the dimensionless temperature $\tilde{T}_{free}$, where,
\begin{eqnarray}
\tilde{T}_{free} \sim N^{-\frac{5-p}{(3-p)^2}} .
\end{eqnarray}
Now recall the strong coupling regime is valid only for $1 \ll \lambda_{eff}^{\frac{1}{5-p}} = \tilde{T}^{-(3-p)/(5-p)}$, otherwise the non-leading terms in the moduli action become important (being dual to $\alpha'$ effects becoming important). Firstly we must check whether the temperature $\tilde{T}_{free}$ is in the allowed range where the strong coupling estimate is valid, which requires,
\begin{eqnarray}
1 \ll \tilde{T}_{free}^{-\frac{3-p}{5-p}} = N^{\frac{1}{3-p}}
\end{eqnarray}
and since we are considering large $N$, we see the free moduli regime only coincides with our strong coupling behaviour in equation \eqref{eq:gravDp} for $p < 3$. Then going to temperatures $T \lesssim \tilde{T}_{free}$, so to stronger coupling, assuming the D$p$-brane behaviour is not modified, then 
the free moduli behaviour will dominate the partition function.
It will be interesting to note that for $p=2$ in fact a new strong coupling transition occurs before this weak thermal scale is reached, due to monopole effects, which changes the strong coupling behaviour to a conformal phase $\epsilon \sim N^{3/2} T^3$, which therefore always dominates the weak thermal behaviour. However for $p=0$ and $p=1$ at sufficiently low temperatures one does indeed reach this weak thermal scale. 
As we shall discuss, for $p=1$ this results in the free orbifold CFT behaviour. 
For $p=0$, the status of this weakly coupled thermal behaviour is less clear due to IR divergences, but would 
represent a weakly interacting gas of D0-branes. However, in fact there is another strongly coupled phase of the theory which becomes relevant at low temperature, also dominating the black D$0$-brane behaviour of \eqref{eq:gravDp}. This has an M-theory dual and we shall discuss it shortly.

%------------------------
\subsection{$p=1$: Free orbifold CFT}
\label{sec:D1}
%------------------------

For $p=1$ at strongly coupled low temperatures $\tilde{T} <  \tilde{T}_{D1} \sim { N^{-\frac{2}{3} } }$ a IIB S-duality on the black D1-brane gives a controlled description in terms of a dual black F1-string solution, with the same thermodynamic behaviour. However at sufficiently low temperature this solution also breaks down, and the theory is expected to transition to the free orbifold CFT behaviour 
\cite{Harvey:1995tg,Bershadsky:1995vm,Dijkgraaf:1997vv}. This occurs at $\tilde{T} \sim  \tilde{T}_{free} \sim 1/N$, and for lower temperatures a thermodynamics,
\begin{eqnarray}
\label{eq:freeorb}
\epsilon \sim N T^2
\end{eqnarray}
is expected. In fact this free orbifold CFT is nothing other than the weakly coupled moduli theory. 
For $p=1$ the gauge fields have no local dynamics, and at weak coupling the bosonic degrees of freedom are free conformal scalar fields. 
The subtlety we noted at footnote \ref{ftnt-symmetric} is that the target space metric for the scalar fields is not that of $(R^{8})^N$, but rather it is the symmetric product $S_N R^8$, due to the branes being identical. As discussed, in the gauge theory derivation of the moduli action this orbifolding occurs due to the residual gauge transformations that permute the diagonal entries in the scalar matrices.

Thus the moduli theory for $p=1$ naively extrapolated to strong coupling for $\tilde{T} \ll 1$ correctly yields the thermal behaviour which in the regime $\tilde{T}_{D1} < \tilde{T}$ is described by the black D$1$-brane, and for $ \tilde{T}_{free} < \tilde{T} < \tilde{T}_{D1}$ is described by an S-dual black F$1$-brane.  However at  $ \tilde{T} \sim  \tilde{T}_{free}$ the weakly coupled moduli behaviour dominates that of the virialised phase, and this is precisely the free orbifold CFT.

%------------------------
\subsection{$p=2$: M2 brane transition on the 11-circle}
\label{sec:D2}
%------------------------

Perhaps the most elegant very strong coupling example is that for $p=2$. The black M2-brane dual description valid where the D$2$-brane description breaks down at the strongly coupled temperature scale $\tilde{T} <  \tilde{T}_{D2} \sim { N^{-\frac{6}{5} } }$ is the dimensional oxidation of the D$2$, and the M$2$ brane is smeared over the M-theory 11-circle. As the temperature further decreases gravity analysis of the smeared M2 brane implies a Gregory-Laflamme transition will occur to a black M2 localised on the 11-circle at a dimensionless temperature $\tilde{T}_{11} \sim { N^{-\frac{3}{2} } }$. For $\tilde{T} \ll \tilde{T}_{11}$ the 
gravity dual 
will have a new thermodynamics, that of a localised black M2 brane, where in our Yang-Mills variables, 
\begin{eqnarray}
\label{eq:M2behaviour}
\epsilon=\frac{16 \pi^4}{81\sqrt{3 \Omega_7}}  N^{3/2} T^3
\end{eqnarray}
and we note the conformal behaviour is governed only by the temperature, and not the coupling $\lambda_2$.

Now let us consider the $(2+1)$-dimensional SYM theory. It is known that for $SU(2)$ gauge group in $(2+1)$-dimensions, there are monopole corrections to the one-loop action going as \cite{Polchinski:1997pz,Dorey:1997tr};
\begin{eqnarray}
 S_{\text{one-loop}}^{\text{3dSYM}} \sim \int d\tau d^2x \frac{2\pi}{\Omega_6} \frac{ | \partial_\mu \phi |^4 }{  | \phi |^{5} } \left( 1 + e^{- \frac{\phi - i \phi^8 }{g_{YM}^2} } + \ldots \right).
 \end{eqnarray}
 Here $\phi^8$  is the dual of the three-dimensional gauge field, which is a periodic variable $\phi^8 \sim \phi^8+ 2 \pi g_{YM}^2$. 
In \cite{Polchinski:1997pz,Dorey:1997tr} it was argued that higher instanton corrections re-sum 
so that the 4 derivative term is multiplied by the harmonic function associated to the transverse brane space in the 11-d M-theory, with the dual gauge field $\phi^8$ playing the role of the 11-coordinate. Such corrections can be generalised to $SU(N)$~\cite{Fraser:1997xi}, which we 
expect
 yields,
\begin{eqnarray}
\label{eq:eff-monopole}
 S_{\text{one-loop}}^{\text{3dSYM}} \sim \int d\tau d^2x \sum_{a<b} \frac{2\pi}{\Omega_6} \frac{ \left| \partial_\mu \left( \vec{\phi}_a -  \vec{\phi}_b \right) \right|^4 }{  | \vec{\phi}_{a} - \vec{\phi}_{b} |^{5} } \left( 1 + e^{- \frac{ \left| \phi_a - i \phi^8_b \right| }{g_{YM}^2} } + \ldots \right).
 \end{eqnarray}
We may deduce that this correction term becomes important when $\left| \phi_a - i \phi^8_b \right| \sim g_{YM}^2$, and hence, using our estimates, 
\begin{eqnarray}
1 \sim \frac{1}{g_{YM}^2} \chi \sim \frac{N}{\lambda_2} \left(\pi \lambda_2 T^2 \right)^{\frac{1}{3}} \quad \implies \quad \tilde{T} = \frac{1}{\lambda_2} T \sim N^{ -\frac{3}{2} }
\end{eqnarray}
where in the last relation we have dropped geometric factors.
Hence from the moduli theory we can see that the estimates from the leading theory, giving D$2$-brane behaviour, will be modified precisely at the scale $\tilde{T}_{11}$, in exact agreement with the gravity analysis. 

In the case $SU(2)$ the instanton sum can remarkably be computed to yield the 4-derivative term above controlled by a harmonic function in the transverse space to the D$p$-brane in 11-d rather than in 10-d, with the dual gauge field $\phi^8$ playing the role of the extra 11-th dimension \cite{Dorey:1997tr,Itzhaki:1998dd}.
Thus, our expectation is that for $SU(N)$ a similar re-summation may be made to yield
the leading one-loop contribution,
\begin{align}
\label{eq:monopole-full}
 S_{\text{one-loop}}^{\text{3dSYM}} \sim \int d\tau d^2x 
 \sum_{n\in\bbZ}
 \sum_{a<b} \frac{(2\pi)^8 l_{\text{pl}}^9}{\Omega_7} \frac{ \left| \frac{1}{(2\pi)^2l^3_{\text{pl}}} \left\{  \left( \partial \vec{X}_a - \partial  \vec{X}_b \right)^2 + \left(\partial X^8_a-\partial X^8_b\right)^2 \right\} \right|^2 }{  \left| (\vec{X}_{a} - \vec{X}_{b})^2 + (X^8_a-X^8_b +2\pi n R )^2 \right|^{3} } 
\end{align}
where we have suppressed some index structure, and have introduced the radius of the M-theory circle, $R$, and eleven-dimensional Planck length $l_{\text{pl}}$ as,
\begin{align}
X^I_a = 2 \pi \alpha' \phi^I_a \; , \qquad R=g_s \sqrt{\alpha'} \, , \qquad 
l_{\text{pl}}^3=g_s \alpha'^{\frac32}
\label{eq:M-relations}
\end{align}
where the index $I = 1, \ldots, 8$, as the gauge field has been dualised to the scalar $X^8_a$.
This one-loop action, together with the classical moduli  kinetic terms, gives the dynamics of nearly parallel M2-branes transverse to the M-theory 11-circle and weakly interacting gravitationally, and may be derived from 11-d supergravity  following \cite{Morita:2013wfa}. It may also be derived from the moduli dynamics of ABJM theory with $k=1$ \cite{Morita:2013wla}. 

For temperatures $\tilde{T} \ll N^{-3/2}$ the estimate for the scalars $\vec{X}_a$ will be $| \vec{X}_a | \ll R$ and hence only the $n=0$ term in the sum contributes. Then the leading action in the new low temperature strongly coupled phase up to one loop becomes,
\ba
 S_{\text{moduli}}^{\text{3dSYM}} &\sim& \int d\tau d^2x 
 \sum_{a} \frac{1}{(2\pi)^2 l_{\text{pl}}^3} \left(\partial X_a^{I} \right)^2
 +
 \sum_{a<b} \frac{(2\pi)^4 l_{\text{pl}}^3}{\Omega_7} \frac{   \left( \partial {X}^I_a - \partial  {X}^I_b \right)^4  }{  \left| {X}^I_{a} - {X}^I_{b} \right|^{6} } 
\ea
which now 
gives the weakly coupled dynamics of 
M2-branes in a flat transverse space, so localised on the M-theory circle \cite{Morita:2013wfa}. Our estimates will then give precisely the correct thermodynamic behaviour \eqref{eq:M2behaviour} that agrees with the localised black M2-brane. 
As noted above, this thermodynamics with $\epsilon \sim N^{3/2} T^3$ is never dominated by the weakly coupled thermal moduli behaviour $\epsilon \sim N T^3$ at large $N$.

%------------------------
\subsection{$p=0$: D0-brane bound states}
\label{sec:D0}
%------------------------

Unlike the cases for $p=1$ and $p=2$ above, the case $p=0$ again has a transition to a new 
strongly coupled
behaviour at very low temperature, but our understanding of it from the moduli perspective is considerably less clear. 

In the case $p=0$ then going to strongly coupled  low temperatures  $\tilde{T} <  \tilde{T}_{D0} \sim { N^{-\frac{10}{21} } }$ we must lift the black D$0$-brane solution to M-theory. This yields a boosted black string wrapping the M-theory circle. Going to lower temperatures still, gravity predicts that this black string will exhibit a Gregory-Laflamme transition on the circle, the gravity analysis indicating the transition occurs at $\tilde{T} \sim \tilde{T}_{11} \sim N^{-\frac{5}{9}}$. For $\tilde{T} \ll \tilde{T}_{11}$ then gravity predicts the thermodynamics to be that of an 11-d black hole boosted on the M-theory circle, which has a different behaviour,
\begin{eqnarray}
\label{eq:locMbh}
\epsilon \sim \pi^{\frac{16}{7}}N^{\frac{12}{7}} \tilde{T}^\frac{16}{7} \lambda_0^{\frac{1}{3}} .
\end{eqnarray}
Note that the transition temperature $\tilde{T}_{11}$ is precisely the temperature at which the entropy of the theory is $S \sim N$, or the energy is $\epsilon \sim N T$, which is the free scale, $\tilde{T}_{free}$, discussed above.

In fact reproducing this M-theory black hole behaviour is precisely the first context that the strong coupling estimates using the virial theorem where first employed by \cite{Horowitz:1997fr, Li:1998ci}. The key observation of \cite{Horowitz:1997fr} is that it is natural that the moduli theory accounts for an entropy which goes as $\sim N$. However, the method of estimation due to \cite{Li:1998ci} that reproduces the behaviour \eqref{eq:locMbh} is very different to ours. 
The claim is that at the transition point, the D0-branes and hence the moduli fields clump together 
into (zero energy) bound states. 
Thus one cannot consider the thermal behaviour of the moduli, but rather the thermal behaviour of these composite objects. In the Appendix \ref{app-D0} we review the arguments of \cite{Li:1998ci} using a language more familiar to our work here. 
However understanding the dynamics of the bound states of moduli for the $p=0$ BFSS matrix model \cite{Banks:1996vh} is an outstanding problem, and further study to elucidate their thermodynamics.

%------------------------
\subsection{$p=3$: S-duality}
\label{sec:D3}
%------------------------

From our estimates for the case $p=3$ the moduli theory correctly predicts the behaviour of the black D$3$-brane and also confirms that it breaks down for $\lambda_3 \lesssim 1$ where the non-leading terms become relevant. Indeed for $\lambda_3 \ll 1$ the SYM should become perturbative.
However, since $(3+1)$-SYM is expected to be invariant under the S-duality $\lambda_3 \to \tilde{\lambda}_3 = (2 \pi)^2 N^2 / \lambda_3$, 
then going to very strong coupling $N^2 \ll \lambda_3$ should be equivalent to weak coupling in the S-dual theory where $\tilde{\lambda}_3 \ll 1$.
Thus we expect the strongly coupled black D$3$-brane behaviour will only apply in the coupling range $1 \ll \lambda_3 \ll N^2$.
In the gravity dual the black D$3$ brane solution breaks down at strong coupling $\lambda_3 \sim N$ due to string coupling corrections. However, at stronger coupling still, control is restored by bulk S-duality, where the S-dual description is again a black D$3$-brane with the same thermodynamic behaviour, and indeed going to very strong coupling $\lambda_3 \sim N^2 $  this S-dual D$3$-brane description breaks down due to  $\tilde{\alpha}'=g_s \alpha'$ corrections.

The leading moduli theory (\ref{moduli-string-frame}) for $p=3$ is invariant under S-duality after the field redefinition,
\begin{eqnarray}
A^M_a \to \tilde{A}^M_a = \frac{2 \pi N}{\lambda_3} A^M_a \,.
\end{eqnarray}
However, our corrections  to this leading theory, such as the thermal one in equation \eqref{eq:thermal}, are not S-duality invariant. Since the full SYM is S-duality invariant, this implies there are other non-perturbative corrections.
For example, this thermal correction is proportional to $e^{-\beta |\vec{\phi}_{ab}|}$ and is relevant in the weak coupling regime $\lambda_3 \ll 1$.
Then S-duality predicts that corresponding terms proportional to the same exponential factor with ${\phi}^I \to \tilde{\phi}^I$ must exist, and hence be proportional to,
\begin{eqnarray}
e^{- \frac{2 \pi N }{\lambda_3} \beta |\vec{\phi}_{ab}|} \sim e^{- \frac{1}{g_{YM}^2} 2 \pi \beta |\vec{\phi}_{ab}|} \,.
\end{eqnarray}
The form of the exponent  going as $\sim 1/g_{YM}^2$ is obviously  suggestive that such terms arise from instanton corrections, and it would be very interesting to see how such terms arise by direct computation.  From our estimates in equation \eqref{eq:chi/T} we see $\beta |\vec{\phi}_{ab}| \sim \sqrt{\lambda_3}$ and thus such non-perturbative corrections to the leading moduli action would indeed become relevant when $\lambda_3 \sim N^2$, and signal the breakdown of the leading moduli theory.

%------------------------
\subsection{$p=4,5,6$}
\label{sec:Dothers}
%------------------------

In the cases $p=4,5,6$ going to strong coupling temperatures beyond $\tilde{T}_{Dp}$ the black D$p$-brane description breaks down. However, a new gravity description that is S-dual or an M-theory lift indicates that the thermodynamics remains unchanged. Indeed we know of no corrections to the leading moduli theory that change its behaviour. 
For example,  potential dominance by the free moduli thermal behaviour discussed earlier only occurs for $p<3$. 
Hence our moduli theory agrees with the new S-dual or M-theory gravity predictions in this very strongly coupled regime.

Let us briefly consider why our estimates in the SYM moduli theory continue to reproduce the correct thermodynamic behaviour, even when the D$p$-brane dual breaks down.
In \cite{Morita:2013wfa} the moduli dynamics of $N$ nearly parallel branes weakly interacting gravitationally was discussed. As we have reviewed, the SYM moduli theory precisely realises this dynamics for D$p$-branes. However, in these cases $p=4,5,6$ it also precisely realises the dynamics of the other dual branes. For $p=4$ it reproduces the moduli dynamics of the lifted M$5$-brane after it has been dimensionally reduced on the 11-circle. Similarly, for $p=5$, the D$5$ moduli dynamics is just the same as the S-dual NS$5$ brane dynamics. Hence the estimates of the M$5$ and NS$5$ thermodynamics simply reproduce the same result as the estimates for the D$4$ and D$5$-brane thermodynamics, and thus our SYM moduli estimates. This is reviewed in more detail in Appendix \ref{app:fullstructure}.

%------------------------
\subsection{Summary}
\label{sec:summary}
%------------------------

In summary, the moduli theory correctly predicts the thermodynamic behaviour of dual black D$p$-branes, but also correctly predicts that this thermodynamic behaviour continues to stronger coupling beyond the applicability of the gravity dual. 
For all cases, except $p=0$, the SYM and moduli theory, with natural assumptions for non-perturbative corrections, allows one to understand when this particular thermodynamic behaviour changes to that of a new phase.
 For $p=0$, there is some understanding of why our estimates break down at the transition point predicted in M-theory.

As an example application of these ideas in Appendix \ref{app:fullstructure} we have described the full phase diagram of maximally supersymmetric $(p+1)$-Yang-Mills on a spatial circle. This combines the various strong coupling phenomena above together with the Gregory-Laflamme transitions on the circle. 
The reader will see that the dual gravity description is extremely complicated,
involving D$p$, M$2$, M$5$, F$1$, NS$5$, gravitational waves and KK monopoles, and yet the predictions from the moduli theory simply involve the estimates above, and knowledge of the corrections due to the circle or 
very
strong coupling.

%-----------------------------------------------------------------
%
\section{Outlook}
\label{sec:outlook}
%
%----------------------- ------------------------------------------

We have discussed how we may use the theory of the SYM moduli, which are computed in a weak coupling limit, to make estimates of the strong coupling behaviour of the thermal SYM, 
by applying our previous ideas \cite{Morita:2013wfa} to the case of SYM.
Remarkably, such strong coupling extrapolation yields precisely the correct parametric behaviour for the thermodynamics, including certain geometric factors such as $\pi$'s and sphere volumes. Presumably this results from non-renormalisation theorems associated to the maximal supersymmetry, and is related to the well known $4/3$ factor in the free energy density going from the perturbative regime to strong coupling having a simple form, when in principle it could have had a complicated transcendental structure. Furthermore, as was known previously \cite{Wiseman:2013cda}, the moduli theory is self consistent in that it predicts its own breakdown when the perturbative SYM regime is reached.

We have emphasised that beyond estimating the thermodynamics, the moduli picture also allows us to predict the structure  of thermal vevs of large classes of local operators. Among these, following the arguments of \cite{Morita:2013wfa}, it also provides an identification of the thermal vev of the scalars to the dual horizon size, where we expect this identification to hold up to a numerical constant of proportionality which does not contain any factors of $\pi$.
 It would obviously be very interesting to understand whether these predictions can be tested by a dual gravity analysis.

Something we have emphasised is that the regime where the strong coupling estimates predict the SYM behaviour is very complicated to describe from a dual gravity perspective. In particular for even $p$ the IIA dual gravity black D$p$-brane description breaks down at very strong coupling and an M-theory description is required. For odd $p$ the IIB gravity breaks down and an S-dual description in terms of F1, dual D3 or NS5 branes is required. However, the moduli theory simply applies in all these regimes with no modification. 
As such it can be seen to provide a unified view of the various phases of the thermodynamics of D$p$, M$2$, M$5$, F$1$ and NS$5$ branes.

We have extended discussion of the moduli theory to include a compact spatial circle. The leading moduli theory is computed at weak coupling, and when estimates are made to extrapolate to strong coupling, these correctly account for the phase transitions on the circle, associated to gauge theory confinement/deconfinement transitions and in the dual gravity, Gregory-Laflamme behaviour. 
In the case of $p=2$, we have also shown how a strong coupling transition occurs, again changing the thermal behaviour, that is correctly accounted for in the moduli approach due to the monopole correction of \cite{Polchinski:1997pz}. In this case the dual interpretation  is a Gregory-Laflamme behaviour on the M-theory circle.
We have discussed other transitions that are predicted by the dual string/M theory to occur at very strong coupling, and have argued that these may all be simply understood within the moduli picture.

As an illustration of these ideas in the Appendix \ref{app:fullstructure} we have provided a full discussion of the thermal phase diagram for SYM with one compact circle. The purpose of this is to demonstrate that a simple application of the estimates, and understanding of possible strong coupling corrections to the moduli theory, can explain a range of behaviours that are very complicated to deduce using dual holographic arguments, requiring many different dual gravity descriptions to cover the full phase diagram.

In this work we have treated the moduli as a predictive tool. 
However, one might wonder whether the moduli theory, restricted to its leading terms (to all derivative order), makes sense in its own right when it is strongly coupled. This point of view was raised in our previous work \cite{Morita:2013wfa}. We stress this point of view is not necessary when considering the moduli as a tool to understand SYM behaviour, but it is clearly an interesting question. 
In particular, it would be interesting to understand the significance of the leading terms of the moduli theory at strong coupling through the scale transformation in equation \eqref{eq:scale}. 

In conclusion we have argued that the moduli theory provides a simple and predictive tool to analyse strongly coupled thermal SYM behaviour in diverse dimensions, including cases where some of these dimensions are compact. Of course it does not give precise answers. However unlike usual perturbation theory, it can be computed in a weak coupling regime and then simply extrapolated to strong coupling to yield the correct behaviour,  making strong predictions for the behaviour of the thermal vevs of many classes of local operators, including the energy density. 
Our hope is that such a simple and unified gauge theory description of the salient features of the strongly coupled SYM thermal phase diagram, and prediction for local operators in these phases, 
will help inform future approaches to directly solve the thermal SYM, including numerical studies.
In particular, it allows a straightforward and intuitive understanding of the thermal behaviour, which is usually obscured in direct approaches.

%-----------------------------------------------------------------
%
\section*{Acknowledgements}
%
%-----------------------------------------------------------------

We would like to thank Kostas Skenderis, David Tong and Arkady Tseyltin for useful discussions.

\appendix

%-----------------------------------------------------------------
%
\section{Summary of thermal phases for $(p+1)$-SYM on a circle}
\label{app:fullstructure}
%
%-----------------------------------------------------------------

\begin{figure}
\begin{tabular}{cc}
\begin{minipage}{0.5\hsize}
\begin{center}
\includegraphics{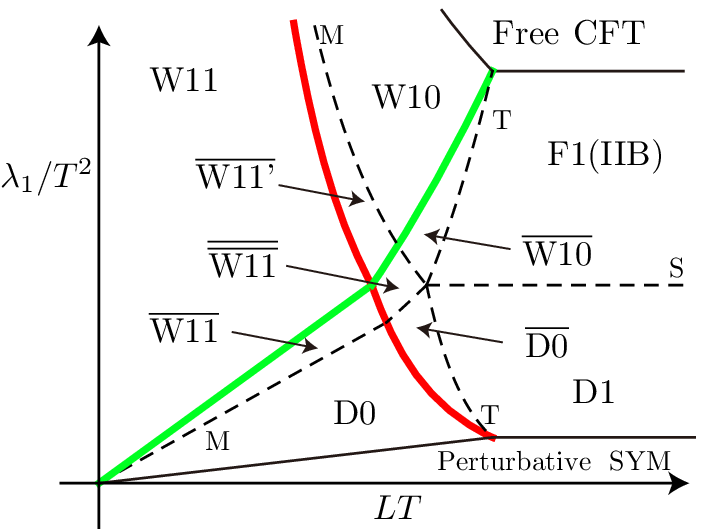}
\end{center}
\end{minipage}
\begin{minipage}{0.5\hsize}
\begin{center}
\includegraphics{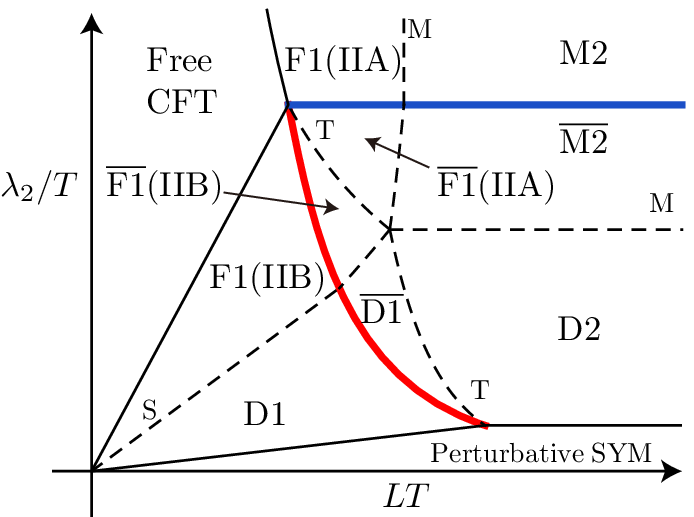}
\end{center}
\end{minipage}
\end{tabular}
\caption{Phase structure of the 2d SYM (LEFT) and  3d SYM (RIGHT) on a circle with the radius $L$.
The red lines denote the GL transitions.
The green lines denote the phase transition associated with the zero energy bound state in the 0-brane.
The blue lines are the phase transition related to the monopole potential of the D2-brane.
The dotted lines denote the borders where the description of the brane solution in the supergravity becomes subtle and we need to use the dualities (S or T-duality, or M-theory lift).
The bar `$\overline{~~~}$' denotes the smeared objects.
}
\label{Fig-D0} 
\end{figure}

We have studied the corrections to the leading moduli theory (\ref{moduli-string-frame}) from  compactification on a spatial circle in Section \ref{sec:circle} and from very strong coupling effects in Section \ref{sec:corrections}.
In order to illustrate these ideas we consider the $p+1$-dimensional SYM theories ($p=1,\dots,6$) compactified on a circle with radius $L$.
The results are summarised in Fig.~\ref{Fig-D0}, \ref{Fig-D2} and \ref{Fig-D4}.
Importantly these results are qualitatively in agreement with supergravity, including in the M-theory regime.

The SYM leading moduli theory describes the dynamics of weakly interacting nearly parallel D$p$-branes. 
We will term this dynamical theory the moduli theory of the branes, as in \cite{Morita:2013wfa}.
We will see that this moduli theory is mapped to the theory describing the dynamics of different branes related by dualities as argued in Section \ref{sec:circle} and  \ref{sec:corrections}.
Once we obtain this moduli action for the branes, our strong coupling estimates predict its thermodynamics will reproduce that of the corresponding black branes \cite{Morita:2013wfa}.

Since the phase structure of the SYM theories are similar, we here will show the details of the thermodynamics of the two-dimensional SYM theory (LEFT of Fig.~\ref{Fig-D0}), and we then will discuss  the other SYM theories briefly.

%-----------------------------------------------------------------
%
\subsection{Two-dimensional SYM theory on a circle}
\label{sec:D0-phase}
%
%-----------------------------------------------------------------

In this section we consider the phase structure of the two-dimensional SYM theory on a spatial circle with radius $L$.

\paragraph{Perturbative SYM}
If the coupling constant is small, $\lambda_1/T^2 \ll 1$, we can evaluate the thermodynamics perturbatively as argued in Section \ref{sec:weak}.
Because of the compactification, there is a phase transition related to the GL transition that occurs even in this region \cite{Aharony:2004ig}.
We have omitted this phase transition in Fig.\ref{Fig-D0}, since it is beyond the scope of this paper.
See \cite{Mandal:2009vz, Mandal:2011hb} for the analytic calculation of this transition.

\paragraph{Black D1 brane}
At strong coupling ($\lambda_1/T^2 \gg 1$), the effective theory of this system is given by the leading moduli theory (\ref{moduli-string-frame})
\ba
S_{\text{moduli}}^{\text{2dSYM}}  &= & \frac{1}{2\pi g_s \alpha'} \int d \tau \int_0^{2\pi L} d x^1  \sum_{a=1}^N  \frac{1}{2} (\partial X_a^I)^2 
\nt &&
 - \frac{(2\pi)^7 g_s^2 \alpha'^{4}}{\Omega_7}  \int d\tau \int_0^{2\pi L} d x^1  \sum_{a<b}^{N}  \left( \frac{1}{2\pi g_s \alpha'} \right)^2
 \frac{  \left( \partial \vec{X}_{ab}   \right)^4  }{ \left| \vec{X}_{ab} \right|^{6} } + \cdots   \,.
 \label{eq:D1-moduli-example}
\ea
Here we have ignored the contribution of the gauge field, that was argued to appear as in (\ref{eq:int-compact}), by assuming $L \phi \gg 1$.
This effective action is the same as the moduli action of the D1-brane (\ref{Dp-moduli-action}) wrapping the compact circle $L$.
This D1-brane moduli action reproduces the black D1-brane thermodynamics from the general argument in Section \ref{sec:SYM}.
This phase corresponds to the region labelled by `D1' in Fig.~\ref{Fig-D0}.

\paragraph{S-duality and F1-strings}
We can rewrite the moduli action (\ref{eq:D1-moduli-example}) by using the parameters $g^{\text{F1}}_s \equiv 1/g_s$ and $\alpha'^{\text{F1}} \equiv g_s \alpha' $, obtaining
\ba
S_{\text{moduli}}^{\text{2dSYM}}  &= & \frac{1}{2\pi \alpha'^{\text{F1}}} \int d \tau \int_0^{2\pi L} d x^1  \sum_{a=1}^N  \frac{1}{2} (\partial X_a^I)^2
\nt && 
 - \frac{(2\pi)^7 \left(g^{\text{F1}}_s\right)^2  \left( \alpha'^{\text{F1} }\right)^4}{\Omega_7}  \int d\tau \int_0^{2\pi L} d x^1  \sum_{a<b}^{N}  \left( \frac{1}{2\pi \alpha'^{\text{F1} }} \right)^2
 \frac{  \left( \partial \vec{X}_{ab}   \right)^4  }{ \left| \vec{X}_{ab} \right|^{6} } + \cdots   \,.
 \label{eq:F1-moduli-example}
\ea
This action represents gravitationally interacting F1-strings, whose tension is $1/\alpha'^{\text{F1}}$.
\footnote{For instance, this can be seen by comparing this action to the general formula for the moduli action presented in \eqref{moduli-string-frame}.}
In IIB supergravity, if the coupling is large ($\lambda_1/T^2 \gg N^{\frac43} $), the black D1-brane solution is not reliable and we should use  $S$-duality and instead consider the black F1-brane solution.
Correspondingly the moduli effective theory of the D1-brane 
may provide the microscopic description of the black F1-brane through (\ref{eq:F1-moduli-example}).
This region is labelled by `F1' in Fig.~\ref{Fig-D0}.

\paragraph{Free orbifold CFT}
At very strong coupling the system will flow to the orbifold CFT, as argued in Section \ref{sec:D1}.
There, the energy is given by
\begin{align}
E \sim NT^2 L \,.
\end{align}
Because of the compactification, there may be some additional phase structure within this conformal phase, but it is beyond the scope of this article.

\paragraph{GL transition to localised D0-brane}
So far we have considered strong coupling effects in the 2d SYM theory.
We now turn our attention to finite $L$ effects.
As argued in Section \ref{sec-Dp-winding-circle}, we can map the D1 moduli action (\ref{eq:D1-moduli-example}) to the D0 brane moduli action  \footnote{
We obtained this equation from (\ref{eq:moduli-T-dual}) and the dimensional reduction of the compact space.
We note that the dimensional reduction would not affect the low energy thermodynamics which we are interested in.
}
\begin{align}
S_{\text{moduli}}^{\text{2dSYM}}  = & \frac{1}{g_s' \alpha'^{1/2}} \int d \tau   \sum_{a=1}^N  \frac{1}{2} (\partial_\tau X_a^I)^2 \nonumber \\
& -  \frac{ (2\pi)^{7} g_s'^2 \alpha'^4}{8\cdot 7  \Omega_{8}}  \int d\tau  \sum_{a<b}^{N}  
 \sum_{n \in \mathbb{Z} }
 \frac{ \left(\frac{1}{g_s'\alpha'^{1/2}}\right)^2 \left( \partial_\tau \vec{X}_{ab}   \right)^4  }{ \left| \bigl(\vec{X}_{ab}\bigr)^2+\left(X^1_{ab}+2 \pi n L'  \right)^2 \right|^{7/2} } + \cdots   \,.
 \label{eq:D0-moduli-example}
\end{align}
Here we have taken $X^1$ as the circle direction, and used the T-duality relation 
$L' \equiv \alpha'/ L$ and $g_s' \equiv g_s \sqrt{\alpha'}/L $.
In this model, the phase transition corresponding the GL transition will occur around $X\sim L' \equiv \alpha'/L$.
 In terms of the temperature, the transition occurs at 
\begin{align}
T_{\text{GL}} \sim \frac{1}{\pi\lambda_0^{\frac12}} \left( \frac{L'}{\alpha'} \right)^{\frac52}
\sim\frac{1}{ \lambda_1^\frac12 L^2}\,.
\end{align}
This is the red line in Fig.~\ref{Fig-D0}.
If $L$ is smaller than this critical size, the D0-branes are clumped along the $X^1$ circle.
This phase is labelled by `D0' in Fig.~\ref{Fig-D0}.
The thermodynamics of this model is given by (\ref{eq:gravDp}) with $p=0$ and $\lambda_0=\lambda_1/2 \pi L$.

\paragraph{Eleven-dimensional boosted black hole}

Below the critical temperature (\ref{eq:T_c-D-dBH}) with $D=11$ and $ 2\kappa_D^2 =(2\pi)^8 l_{\text{pl}}^9 $, 
the D0-branes may form zero-energy bound states. When this occurs the effective action (\ref{eq:D0-moduli-example}) becomes 
\begin{align}
S_{\text{moduli}}^{\text{D0,bound}}  = & \frac{N}{RM} \int d \tau   \sum_{a=1}^M  \frac{1}{2} (\partial_\tau X_a^I)^2 \nonumber \\
&
 -  \frac{(2\pi)^{7} l_{\text{pl}}^9 }{8\cdot 7 \Omega_{8} R }  \int d\tau  \sum_{a<b}^{M}  
 \sum_{n \in \mathbb{Z} }
 \sum_{\delta p_-=0}^{N/M} \left( \frac{N}{RM} \right)^2 f(\delta p_-)
 \frac{  \left( \partial_\tau \vec{X}_{ab}   \right)^4  }{ \left| \bigl(\vec{X}_{ab}\bigr)^2+\left(X^1_{ab}+2 \pi n L'  \right)^2 \right|^{7/2} } + \cdots   \,.
  \label{eq:D0-moduli-bound-example}
\end{align}
Here we have used the M-theory parameters (\ref{eq:M-relations}) with $g_s \to g_s'=g_s \sqrt{\alpha'}/L $, since we had taken the T-dual.
If $L'$ is large, we can ignore sum of the mirror branes and the thermodynamics is given by (\ref{eq:0-brane-bound-result}) with $D=11$.
The label `W11' in Fig.~\ref{Fig-D0} represents this phase.

\paragraph{Ten-dimensional wave in IIA supergravity}
Starting from the moduli effective action for the F1-string (\ref{eq:F1-moduli-example}) and taking the radius $L$ to be small, we obtain
\begin{align}
S_{\text{moduli}}^{\text{2dSYM}}=  \frac{1}{\tilde{L}} \int d \tau   \sum_{a=1}^N  \frac{1}{2} (\partial X_a^I)^2 
 - \frac{(2\pi)^7\tilde{ g}_s^2  \left( \alpha'^{\text{F1} }\right)^4 }{2\pi\tilde{L}\Omega_7} \int d\tau  \sum_{a<b}^{N}  \left( \frac{1}{\tilde{L}} \right)^2
 \frac{  \left( \partial \vec{X}_{ab}   \right)^4  }{ \left| \vec{X}_{ab} \right|^{6} } + \cdots   \,,
 \label{eq:wave-moduli-example}
\end{align}
via a KK reduction.
Here $\tilde{L}= \alpha'^{\text{F1} }/L$ and $\tilde{ g}_s = g^{\text{F1}}_s \sqrt{\alpha'^{\text{F1} }}/L $.
This action describes $N$ gravitationally interacting particles whose masses are $1/\tilde{L}$ in nine dimensions. This nine-dimensional spacetime results from the KK reduction from ten dimensions on a circle whose radius is $\tilde{L}$.
Since $1/\tilde{L}$ is the KK mass, we interpret the action (\ref{eq:wave-moduli-example}) as the moduli action for the waves which carry the KK momentum along the $S^1_{\tilde{L}}$.
This interpretation is natural because we can obtain the wave from the winding F1-string (\ref{eq:F1-moduli-example}) via T-duality.

The waves can form a zero-energy bound state, and as we shall argue in Appendix \ref{app-D0}, this bound state will dominate at low temperatures.
The effective action for the bound state is given by,
\ba
S_{\text{moduli}}^{\text{D0,bound}} &= & \frac{N}{\tilde{L}M} \int d \tau   \sum_{a=1}^M  \frac{1}{2} (\partial_\tau X_a^I)^2 \nonumber \\
&&
 - \frac{(2\pi)^7\tilde{ g}_s^2  \left( \alpha'^{\text{F1} }\right)^4 }{2\pi\tilde{L}\Omega_7} \int d\tau  \sum_{a<b}^{M}  
 \sum_{\delta p_-=0}^{N/M} \left( \frac{N}{\tilde{L}M} \right)^2 f(\delta p_-)
 \frac{  \left( \partial_\tau \vec{X}_{ab}   \right)^4  }{ \left| \vec{X}_{ab} \right|^{6} } + \cdots   \,.
  \label{eq:wave-bound-moduli-example}
\ea
Through a similar calculation to that of the D0-brane example, 
we can estimate the thermodynamics of this system as 
\begin{align}
X \sim \left( \frac{\tilde{g}^2_s   \left( \alpha'^{\text{F1} }\right)^4 NT }{\pi^3 \tilde{L}} \right)^{\frac16}, \qquad 
E \sim \left( \frac{\pi^3 N^4 \tilde{g}^2_s   \left( \alpha'^{\text{F1} }\right)^4 }{\tilde{L}^4} \right)^{\frac13} T^{\frac73}.
\end{align}
In ten dimensions the wave corresponds to the boosted black string, whilst the wave bound state corresponds to the boosted localised black hole.
The labels $\overline{\text{W10}}$ and W10 in Fig.~\ref{Fig-D0} represent these phases.

Note that the moduli effective actions for the ten-dimensional wave (\ref{eq:wave-bound-moduli-example}) are obtained from the effective action for the D0-brane (\ref{eq:D0-moduli-bound-example}) by taking $L'$ to be small and identifying $L'$ as the new M-theory circle.
\footnote{For this reason, we write the right hand side of the action (\ref{eq:wave-bound-moduli-example}) as $S_{\text{moduli}}^{\text{D0,bound}}$.}
To see this, we may use the following relations
\begin{align}
R= \tilde{L}, \qquad L'= \tilde g_s \sqrt{\alpha'^{\text{F1}}}, \qquad l^3_{\text{pl}}= \tilde g_s \left(\alpha'^{\text{F1}}\right)^{\frac32}.
\end{align}
These relations may be compared with those of \eqref{eq:M-relations}.
One can see that the roles of the M-theory circle, $R$, and the compact circle, $L'$, in the original D0-brane theory (\ref{eq:D0-moduli-example}) have been swapped in the wave theory (\ref{eq:wave-moduli-example}). This is because we have taken an S-duality in obtaining this wave theory, as part of the sequence of dualities (D0 $\xrightarrow{T}$ D1 $\xrightarrow{S}$ F1 $\xrightarrow{T}$ wave).

\subsection{Other SYM theories on a circle}
In this section we briefly explain the phase structure of the $p+1$-dimensional SYM ($p=2, \dots, 6 $) on a circle with radius $L$.
Since the phases are structurally similar to the 2d SYM case, we will only highlight some important features of their respective phase diagrams.

%-----------------------------------------------------------------
%
\subsubsection{3d SYM} 
%
%-----------------------------------------------------------------

The phase diagram for 3d SYM on a circle is shown in Fig.~\ref{Fig-D0} (RIGHT), where four phases appear.
One is the free orbifold CFT, and the others may be labelled by localised black D1-branes (D1), the black D2-branes (D2) and localised black M2-branes (M2). 
Each phase may also be described using other branes, by applying appropriate dualities at the location of the dotted lines in Fig.~\ref{Fig-D0}, just as in Section \ref{sec:D0-phase}.

The phase transitions among these phases are related either to the GL transition, or to  the monopole effect associated with the GL transition along the M-theory circle. These are shown by the red and blue lines in Fig.~\ref{Fig-D0}, respectively.

\paragraph{F1-strings from M2-branes}
In this phase diagram the IIA F1-string regime is connected to the M2-brane regime.
Related to this structure, we can see that the moduli actions for these branes are also related.
The moduli action for the M2-branes appear in the 3d SYM theory as shown in (\ref{eq:monopole-full}). In the case where the 3d SYM theory is compactified on a circle with radius $R$, the M2-brane moduli action becomes\begin{align}
 S_{\text{moduli}}^{\text{M2}} =  &
 \int d\tau dx^1 \int_0^{2\pi R} dx^2 
 \sum_{a} \frac{1}{(2\pi)^2l_{\text{pl}}^3} \left(\partial X_a^{I} \right)^2  +
 \sum_{a<b} \frac{(2\pi)^8l_{\text{pl}}^9}{\Omega_7} \frac{ \left| \frac{1}{(2\pi)^2l^3_{\text{pl}}} \left( \partial \vec{X}_a - \partial  \vec{X}_b \right)^2  \right|^2 }{ (\vec{X}_{a} - \vec{X}_{b})^6} \,.
\end{align}
By taking $R$ small and considering the KK-reduction, we obtain
\begin{align}
 S_{\text{moduli}}^{\text{M2}} \sim  &
 \int d\tau dx^1  
 \sum_{a} \frac{R}{2\pi l_{\text{pl}}^3} \left(\partial X_a^{I} \right)^2  +
 \sum_{a<b} \frac{(2\pi)^7 l_{\text{pl}}^9}{\Omega_7 R} \frac{ \left| \frac{R}{2\pi l^3_{\text{pl}}} \left( \partial \vec{X}_a - \partial  \vec{X}_b \right)^2  \right|^2 }{  (\vec{X}_{a} - \vec{X}_{b})^6 } \,.
\end{align}
By using the relations $l_{\text{pl}}^3/R = \alpha'_{\text{F1}}$ and $l_{\text{pl}}^9/R = g_{s\text{F1}}^2 \alpha_{\text{F1}}'^4=2 \kappa^2_{\text{F1}}/(2\pi)^7$ (\ref{eq:M-relations}), we can identify this action as the moduli action for the F1-strings.
The thermal energies of F1 and M2-branes derived from these moduli actions are related, 
\ba
E_\text{F1}\sim (2\pi)^3 N^\frac32 T^3 g_{s\text{F1}} \sqrt{\alpha'_{\text{F1}}} V_1
= (2\pi)^2 N^\frac32 T^3 (2\pi R V_1)
\sim E_\text{M2}\,.
\ea
This result agrees with the fact that IIA F1-string is obtained from M2-brane through the double dimensional reduction.

%-----------------------------------------------------------------
%
\subsubsection{4d SYM} 
%
%-----------------------------------------------------------------

\begin{figure}
\begin{tabular}{cc}
\begin{minipage}{0.5\hsize}
\begin{center}
\includegraphics{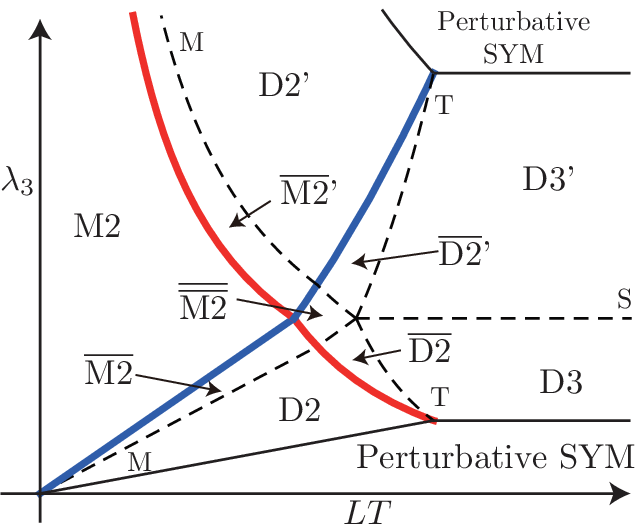}
\end{center}
\end{minipage}
\begin{minipage}{0.5\hsize}
\begin{center}
\includegraphics{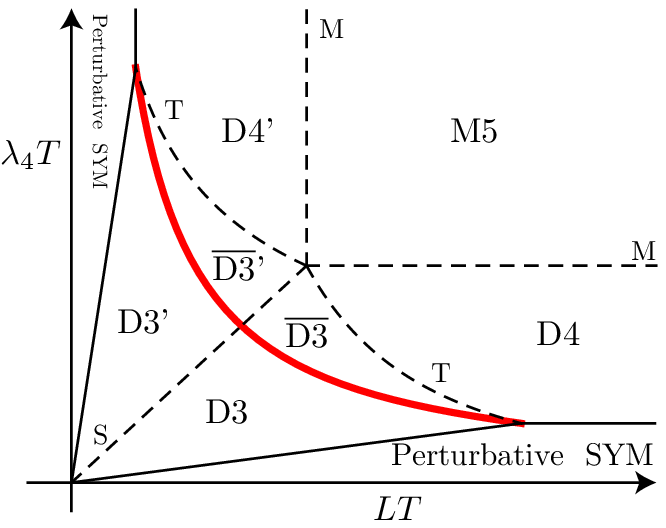}
\end{center}
\end{minipage}
\end{tabular}
\caption{Phase structure of the 4d SYM (LEFT) and 5d SYM (RIGHT) on a circle with the radius $L$.}
\label{Fig-D2} 
\end{figure}

The phase diagram of 4d SYM theory on a circle is shown in Fig.~\ref{Fig-D2} (LEFT), where four phases appear.
These phases are described by the D3-brane moduli theory on a circle
\ba
 S_{\text{moduli}}^{\text{4dSYM}} & \sim  &
 \int d\tau d^2x \int_0^{2\pi L} dx_3 
 \sum_{a} \frac{1}{(2\pi)^3 g_s \alpha'^2} \left(\partial X_a^{I} \right)^2 \nonumber \\
&& +
 \sum_{n,m\in\bbZ}
 \sum_{a<b} \frac{(2\pi)^7 g_s^2\alpha'^4}{\Omega_7} \frac{ \left| \frac{1}{(2\pi)^3g_s\alpha'^2} \left\{  \left( \partial \vec{X}_a - \partial  \vec{X}_b \right)^2 + (2\pi\alpha')^2\left( \partial a_a^3 - \partial  a^3_b \right)^2
 + \left( \partial X_a^8 - \partial  X^8_b \right)^2
  \right\} \right|^2 }{  \left| (\vec{X}_{a} - \vec{X}_{b})^2 + (2\pi\alpha')^2(a^3_a-a^3_b + n /L )^2+ (X^8_a-X^8_b +2\pi m R )^2 \right|^{3} }
  \nt
\ea
where $a^3_a$ are gauge moduli for the winding direction, and the interactions arise as argued in (\ref{sec-Dp-winding-circle}).
$X^8$ is the scalar dual to the components of the gauge field in the three non-compact directions $A_{0,1,2}$, which may appear by regarding the T-duality to the 3d SYM theory and the monopole effect.
As argued in Section \ref{sec:D2},  $X^8$ corresponds to the M-theory direction. 
$R$ is the M-theory radius defined in (\ref{eq:M-relations}) by replacing $g_s$ and $\alpha'$ regarding the T-duality.

The four phases are characterised by the distributions of $a^3$ and $X^8$ along the two circles.

%-----------------------------------------------------------------
%
\subsubsection{5d SYM} 
\label{sec-5dSYM}
%
%-----------------------------------------------------------------

Here we consider the moduli effective action for the five-dimensional SYM theory
\ba
S_{\text{moduli}}^{\text{5dSYM}} \sim \int d \tau d^{4}x \frac{1}{(2\pi)^4 g_s \alpha'^{\frac52}} \sum_a (\partial X_a)^2 
+  \frac{(2\pi)^7 g_s^2 \alpha'^4}{\Omega_4} \sum_{a,b} \frac{ \left( \frac{1}{(2\pi)^4 g_s \alpha'^{\frac52}} (\partial X_{ab})^2\right)^2  }{X_{ab}^{3}}   
+\cdots .
\label{eq:D4-eff}
\ea
At very strong coupling ($\lambda_4 T \gg N^{\frac{2}{3}}$), this action lifts to the M5-brane effective action  
\begin{align}
S_{\text{moduli}}^{\text{M5}} \sim \int d \tau d^{5}x \frac{1}{(2\pi)^5 l_{\text{pl}}^6} \sum_a (\partial Y_a)^2 
+  \frac{(2\pi)^8l_{\text{pl}}^9}{\Omega_4} \sum_{a,b} \frac{ \left( \frac{1}{(2\pi)^5l_{\text{pl}}^6} (\partial Y_{ab})^2\right)^2  }{Y_{ab}^{3}}   
+\cdots
\label{eq:eff-M5}
\end{align}
which is obtained from (\ref{moduli-string-frame}) with $\mu=1/(2\pi)^5l_{\text{pl}}^6$ and $2\kappa^2= (2\pi)^8 l^9_{\text{pl}}$, which are the tension of the M5-brane and the eleven-dimensional gravitational coupling respectively.
The M5-brane moduli action can also be estimated from the six-dimensional (2,0) superconformal field theory by using a non-renormalisation theorem~\cite{Morita:2013wla}.

Indeed, if we compactify one of the longitudinal directions of the M5-brane effective action (\ref{eq:eff-M5}), say $x^5$, on a circle with radius $R$, the moduli fields may be expanded as
\ba
Y^I_a=X^{I(0)}_a+\sum_{n\neq 0} X^{I(n)}_ae^{inx_5/R}
\ea
and, by using the relation (\ref{eq:M-relations}), the M5-brane effective action (\ref{eq:eff-M5}) becomes
\ba
 S^{\text{M}5}_{\text{moduli}} &=& S_0^{\text{M}5} + S_2^{\text{M}5} +\cdots \\
 S_0^{\text{M}5}&\sim& \int d\tau d^4x \sum_a \frac{1}{(2\pi)^4 g_s \alpha'^{\frac52}} \left[(\partial X_a^{(0)})^2
+\sum_{n\neq 0}\left( |\partial X_a^{(n)}|^2 +  \frac{n^2}{R^2} | X_a^{(n)}|^2 \right) \right], \nt
 S_2^{\text{M}5}&\sim& \frac{(2\pi)^7 g_s^2\alpha'^4}{\Omega_4} \int d\tau d^4x \sum_{a<b}
\frac{\left( \frac{1}{(2\pi)^4 g_s \alpha'^{\frac52}} \left(\partial X_{ab}^{(0)}\right)^2\right)^2}{\left(X_{ab}^{(0)}\right)^6}+(\text{terms including KK modes})\,.\nn
\ea
Thus the action for the KK zero-modes agrees with the D4-brane moduli action 
(\ref{eq:D4-eff})\footnote{This holds not just for the scalar moduli fields, but for the gauge fields too, which are related as 
$ F_a^{ij}=F_a^{(0)ij5} = \epsilon_a^{ij 5 klm}  F_a^{(0)klm} $, where
$F_a^{ij}$ $(i,j=0,\ldots, 4$) is the field strength of the U(1) gauge field on the $a$-th D4-brane and $ F_a^{\mu\nu \rho}$ $(\mu,\nu=0,\ldots, 5$) is the field strength of the self-dual two-form field on the $a$-th M5-brane.
}.
It is a challenging question to derive the contributions from the KK non-zero mode in 5d SYM.

The thermal energies of the D4 and M5-branes which are derived from the moduli actions are connected through
\ba
E_{\text{D}4}\sim (2\pi)^4 N^3T^6g_s \sqrt{\alpha'}V_4 
= (2\pi)^3 N^3T^6(2\pi RV_4) \sim E_{\text{M}5}
\ea
where we have used $V_5 = 2\pi R V_4$
as pointed out in \cite{Morita:2013wla}, and we do not expect any phase transition 
between these two solutions.

%-----------------------------------------------------------------
%
\subsubsection{6d SYM} 
\label{sec-6dSYM}
%
%-----------------------------------------------------------------

\begin{figure}
\begin{tabular}{cc}
\begin{minipage}{0.5\hsize}
\begin{center}
\includegraphics{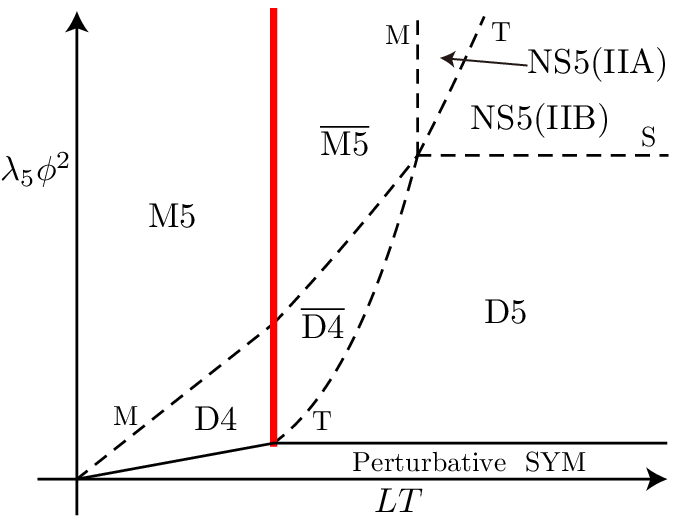}
\end{center}
\end{minipage}
\begin{minipage}{0.5\hsize}
\begin{center}
\includegraphics{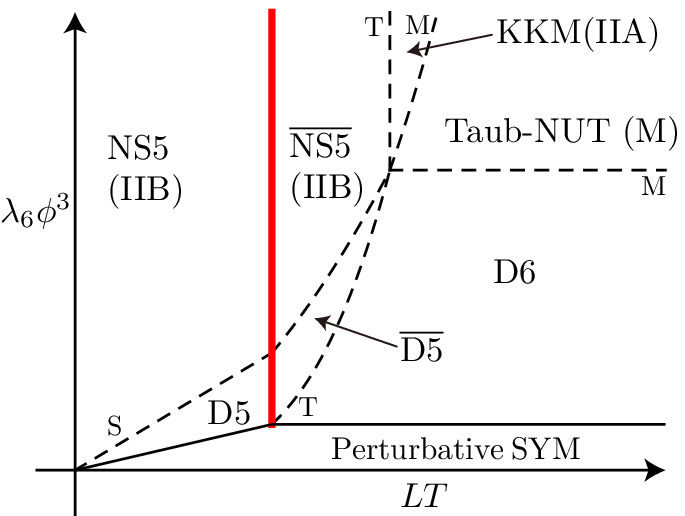}
\end{center}
\end{minipage}
\end{tabular}
\caption{Phase structure of the 6d SYM (LEFT) and 7d SYM (RIGHT) on a circle with the radius $L$.
$\phi$ is the typical value of the scalar moduli which correspond to the location of the horizon in the gravity, which is a free parameter in the Hagedorn regime (e.g. D5).
}
\label{Fig-D4} 
\end{figure}
The phase diagram for 6d SYM on a circle is shown in Fig.~\ref{Fig-D4} (LEFT).
In this phase diagram, two phases appear and a GL transition occurs between them.
One feature of this system is that the D5-brane phase (= smeared D4-brane phase) shows Hagedorn behaviour, as we mentioned in Section \ref{sec-Hagedorn}.
Here the temperature is uniquely fixed as
\begin{align}
T_H \sim \sqrt{\frac{\Omega_3}{\pi\lambda_5}} \sim \frac{1}{\pi}\sqrt{\frac{\Omega_3}{\lambda_4 L}}\,.
\end{align}
Thus we use the scale of the scalar moduli $\phi$ as the parameter to draw the phase diagram instead of the temperature.
For example the GL transition occurs at $X \sim L'=\alpha'/L$, which means that
\begin{align}
\phi= \frac{1}{2 \pi \alpha'} X \sim \frac{L'}{ \alpha'} = \frac{1}{ L}\,.
\end{align}
This corresponds to the red line in the Fig.~\ref{Fig-D4}.

\paragraph{IIB NS5-branes from D5-branes}
Another notable feature of this phase diagram is the appearance of NS5-branes.
The moduli effective action for NS5-branes in IIB superstring theory can be obtained from the D5-brane theory
\ba
S_{\text{moduli}}^{\text{6dSYM}} \sim \int d \tau d^{5}x \frac{1}{(2\pi)^5 g_s \alpha'^{3}} \sum_a (\partial X_a)^2 
+  \frac{(2\pi)^7 g_s^2 \alpha'^4}{\Omega_3} \sum_{a,b} \frac{ \left( \frac{1}{(2\pi)^5 g_s \alpha'^{3}} (\partial X_{ab})^2\right)^2  }{X_{ab}^{2}}   
+\cdots
\label{eq:D5-eff}
\ea
Where we can rewrite $g_s = 1/ \tilde{g}_s$ and $\alpha' = \tilde{g}_s  \tilde{\alpha}'$ to  obtain
\ba
S_{\text{moduli}}^{\text{6dSYM}} \sim \int d \tau d^{5}x \frac{1}{(2\pi)^5\tilde{g}^2_s \tilde{\alpha}'^{3}} \sum_a (\partial X_a)^2 
+  \frac{(2\pi)^7\tilde{g}_s^2 \tilde{\alpha}'^4}{\Omega_3} \sum_{a,b} \frac{ \left( \frac{1}{(2\pi)^5\tilde{g}^2_s \tilde{\alpha}'^{3}} (\partial X_{ab})^2\right)^2  }{X_{ab}^{2}}   
+\cdots
\label{NS5-effective-string-unit}
\ea
where $1/(2\pi)^5 \tilde{g}^2_s \tilde{\alpha}'^{3}$ is the correct tension of the single NS5-brane, and consistent with the moduli effective action (\ref{moduli-string-frame}) independently derived from supergravity.

\paragraph{IIA NS5-branes from IIB NS5-branes} 
The NS5-branes in IIA superstring theory is obtained from the T-dual of the IIB NS5-branes.
Let us consider the IIB NS5-branes winding a circle with a radius $L$.
This situation appears in the 6d SYM on a circle as we showed in \eqref{NS5-effective-string-unit}. See also Fig.~\ref{Fig-D4} (LEFT).
Because of the compact circle, the scalar moduli may be expanded as 
\ba
X_a^I(\tau,x_{1,\ldots,5})
= X_a^{I(0)}(\tau,x_{1,\ldots,4})
+ \sum_{n\neq 0} X_a^{I(n)}(\tau,x_{1,\ldots,4}) e^{inx_5/L}
\ea
Then the moduli effective action (\ref{NS5-effective-string-unit}) becomes 
\ba
S^{\text{6dSYM}}_\text{moduli}
& =& S_0^\text{6dSYM}+S_2^\text{6dSYM}+\cdots \nt
S_0^{\text{6dSYM}} &\sim & \int d\tau d^4x \frac{2\pi L}{(2\pi)^5\tilde{g}_s^2\tilde{\alpha}'^3}\sum_a \left[
 (\partial X_a^{(0)})^2 + \sum_{n\neq 0} |\partial X_a^{(n)}|^2 + \frac{n^2}{L^2} |X_a^{(n)}|^2
\right]
\nt
S_2^{\text{6dSYM}} &\sim & \int d\tau d^4x \frac{(2\pi)^7\tilde{g}_s^2\tilde{\alpha}'^4}{2\pi L\Omega_3}\sum_{a<b}
 \frac{\left(\frac{2\pi L}{(2\pi)^5\tilde{g}_s^2\tilde{\alpha}'^3}(\partial X_{ab}^{(0)})^2\right)^2}{\left(X_{ab}^{(0)}\right)^2} 
+(\text{terms including KK modes})\,.\nt
\ea
We rewrite this action by using the parameters $L' \equiv \tilde{\alpha}'/L$ and  $g_s' \equiv  \tilde{g}_s \sqrt{\tilde{\alpha}'}/L$ and obtain
\ba
S_0^\text{6dSYM} &\sim &
\int d\tau d^4x \frac{2\pi L'}{(2\pi)^5g_s'^2\tilde{\alpha}'^3} \sum_a \left[
 (\partial X_a^{(0)})^2 + \sum_{n\neq 0} |\partial X_a^{(n)}|^2 + 
 \frac{n^2 L'^2}{\tilde{\alpha}'^2}|X_a^{(n)}|^2  \right]
\nt
S_2^\text{6dSYM} &\sim &
\int d\tau d^4x \frac{(2\pi)^7 g_s'^2\tilde{\alpha}'^4}{2\pi L'\Omega_3}\sum_{a<b}\frac{\left(\frac{2\pi L'}{(2\pi)^5 g_s'^2\tilde{\alpha}'^3}(\partial X_{ab}^{(0)})^2\right)^2}{\left(X_{ab}^{(0)}\right)^2}+(\text{terms including KK modes})\,. \nt
\ea
This action can be regarded as IIA NS5-branes winding a circle with radius $L'$.
Thus the 6d SYM theory can describe the moduli dynamics of the IIA NS5-branes.
This description may be natural in the region where the IIA NS5-brane geometry is reliable in the supergravity (the region labelled by NS5(IIA) in Fig.~\ref{Fig-D4} (LEFT)).
Note that the KK modes of IIB NS5-branes are mapped to the winding modes of IIA NS5-branes.\footnote{
In the worldvolume theory of IIA NS5-branes there are additional scalar moduli,
since this theory has an additional scalar field corresponding under T-duality to a component of the 6d Yang-Mills gauge field $A_\sigma$ (where $\sigma$ denotes the T-duality direction) in the IIB NS5-brane worldvolume theory. (See Appendix \ref{app:fields}.) However, we will see our evaluation matches the gravity result without considering these scalar moduli, so we do not discuss them here.}

\paragraph{IIA NS5-branes from M5-branes} 
NS5-branes in type IIA superstring theory may also be obtained from the dimensional reduction of M5-branes which 
unwrap the M-theory circle.
We see this in the M5-brane moduli theory which would appear, for example, in 6d SYM theory on a circle (Fig.~\ref{Fig-D4} (LEFT)).
Because of the compactification, the interaction term of the M5-brane effective action (\ref{eq:eff-M5}) is modified
\begin{align}
\label{eq:M5-compact}
S_2^{\text{M}5} &= \,  \frac{1}{(2\pi)^2 \Omega_{4} l_{\text{pl}}^3 } \int dt d^5x \sum_{a<b}^{} \sum_n  \frac{ 2 \left( \partial_\mu \vec{Y}_{ab} \cdot \partial_\nu \vec{Y}_{ab} \right)^2 -  \left( \partial_\mu \vec{Y}_{ab} \cdot \partial^\mu \vec{Y}_{ab} \right)^2 }{ \left| \left(\vec{X}_{ab} \right)^2+\left(Y_{ab}^{5} +2\pi R n\right)^2    \right|^{\frac32} }    \nonumber \\
& \sim
\,  \frac{(2\pi)^7 g_s^2 \alpha'^4 }{ \Omega_3 } \int dt d^5x \sum_{a<b}^{} \sum_n  \frac{  \left( \frac{1}{ (2\pi)^5 g_s^2\alpha'^3}  \left( \partial_\mu \vec{Y}_{ab}  \right)^2 \right)^2 }{  \left(\vec{X}_{ab} \right)^2 } \left(1+ \sum_{m \ge 1} e^{-\frac{m |Y^5|}{R}} + \cdots \right)  \,.
\end{align}
Here we have used $(\vec{Y})=(\vec{X},Y^{5})$
and \eqref{eq:M-relations}.
This interaction term agrees with the NS5-brane effective action (\ref{NS5-effective-string-unit}) in the limit $R \to 0$.
In addition, this expansion predicts that non-perturbative corrections of the $(2,0)$ supersymmetric theory on the NS5-branes are similar to the monopole corrections in the three-dimensional SYM theory on the D2-branes. 
(In the $(2,0)$ supersymmetric theory on the NS5-branes, $Y^5$ can be identified with the scalar field.)
This non-perturbative effect cannot be ignored when $Y \sim R$ and 
a phase transition may occur there.
Indeed in eleven-dimensional supergravity, the GL transition occurs between the smeared black M5-brane and localised M5-branes around $Y \sim R$.

%-----------------------------------------------------------------
%
\subsubsection{7d SYM} 
\label{sec-7dSYM}
%
%-----------------------------------------------------------------

We now turn to 7d SYM on a circle (Fig.~\ref{Fig-D4} (RIGHT)).
In this phase diagram, two phases appear and a GL transition occurs between them.
This system also shows Hagedorn behaviour in the localised D5 and NS5-brane phase.
One feature of the phase diagram is the appearances of the KK monopole in the IIA superstring and the Taub-NUT in the M-theory.

\paragraph{NS5-branes from D6-branes}

Let us consider the 7d SYM theory (D6-brane theory) compactified on a circle with the radius $L$. The moduli action is given by (\ref{Dp-moduli-action}) with $p=6$.
Since D6-branes can be mapped to NS5-branes through a T-duality along the circle and S-duality in superstring theory, we expect that this moduli action describes the NS5-branes too.
Indeed, if we rewrite the parameters as
\begin{align}
\tilde{\alpha}' \equiv \frac{g_s (\alpha')^{\frac{3}{2}} }{L}, \quad
\tilde{g}_s \equiv \frac{L}{g_s \sqrt{\alpha'}}, \quad \tilde{L}\equiv \frac{\alpha'}{L},
\label{eq-para-NS5}
\end{align}
the moduli action (\ref{moduli-string-frame}) with $p=6$ becomes
\ba
S_{\text{moduli}}^{\text{7dSYM}} & \sim & \int d \tau d^{5}x \frac{1}{(2\pi)^5\tilde{g}_s^2 \tilde{\alpha}'^{3}} \sum_a (\partial X_a)^2 
+ \frac{(2\pi)^7 \tilde{g}_s^2 \tilde{\alpha}'^4}{\Omega_3} \sum_n \sum_{a,b} \frac{ \left( \frac{1}{(2\pi)^5 \tilde{g}_s^2 \tilde{\alpha}'^{3}} (\partial \vec X_{ab})^2\right)^2  }{(\vec X_{ab})^{2}+(X_{ab}^9+ 2\pi n \tilde{L})^{2}}   
+\cdots \nonumber \\
& \sim &\int d \tau d^{5}x \frac{1}{(2\pi)^5\tilde{g}_s^2 \tilde{\alpha}'^{3}} \sum_a (\partial X_a)^2 
+  \frac{(2\pi)^7 \tilde{g}_s^2 \tilde{\alpha}'^4 }{2\pi \tilde{L}\Omega_2} \sum_{a,b} \frac{ \left( \frac{1}{(2\pi)^5 \tilde{g}_s^2 \tilde{\alpha}'^{3}} (\partial \vec X_{ab})^2\right)^2  }{|\vec X_{ab}|} \left(
1+e^{-\frac{|X|}{\tilde{L}}}\cdots 
\right).  \nt
\label{eq-IIBNS5-S1}
\ea
This agrees with the moduli action of NS5-branes (\ref{NS5-effective-string-unit})  localised on a circle with the radius $\tilde{L}$.
Here we have taken the compact direction to be $x^9$ and used \eqref{sec-Dp-winding-circle}.
The second line has been expanded by assuming $|X| \gg \tilde{L}$.

A GL phase transition may occur when $X \sim \tilde{L}$.
Hereafter we focus on the smeared configuration.
Via the virial theorem in the action (\ref{eq-IIBNS5-S1}), the scale of the scalar moduli and thermal energy for the smeared NS5-branes are estimated as
\begin{align}
X_{\overline{\text{NS5}}}^{\text{(IIB)}} \sim \frac{\tilde{L}}{\pi^2 N \alpha' T^2}\,, \qquad 
E_{\overline{\text{NS5}}}^{\text{(IIB)}} \sim \frac{\tilde{L}^2}{\pi^7 N \tilde{g}_s^2\tilde{\alpha}'^5 T^2} V_5\,.
\label{eq:NS5-thermal}
\end{align}
This is the region labelled by `$\overline{\text{NS5}}$ (IIB)' in Fig.~\ref{Fig-D4} (RIGHT).
Note that if we rewrite these results by using the original parameters $\alpha'$, $g_s$ and $L$, we reproduce those of the black D6-branes (\ref{eq:Dp-scale}) and (\ref{eq:gravDp}).

\paragraph{KK monopoles from NS5-branes}
In superstring theory, when $\tilde{L}$ is small, we may perform a T-duality on $x^9$ mapping the NS5-branes to Kaluza-Klein monopoles.
Correspondingly, if we use the parameters
\begin{align}
\tilde{\tilde{g}}_s \equiv \tilde{g}_s \frac{\sqrt{\tilde{\alpha}'}}{ \tilde{L}}, \quad
\tilde{\tilde{L}}\equiv \frac{\tilde{\alpha}'}{\tilde{L}},
\label{eq-para-KKM}
\end{align}
the NS5-brane moduli effective action (\ref{eq-IIBNS5-S1}) becomes 
\ba
S_{\text{moduli}}^{\text{7dSYM}} & \sim & \int d \tau d^{5}x 
\frac{(2\pi\tilde{\tilde{L}})^2}{(2\pi)^7\tilde{\tilde{g}}^2_s \tilde{\alpha}'^{4}} \sum_a (\partial X_a)^2 
\nt &&\qquad
+   \frac{(2\pi)^7\tilde{\tilde{g}}_s^2 \tilde{\alpha}'^4}{2\pi\tilde{ \tilde{L}}\Omega_2}  \sum_{a,b} \frac{ \left( \frac{(2\pi\tilde{\tilde{L}})^2}{(2\pi)^7\tilde{\tilde{g}}^2_s \tilde{\alpha}'^{4}} (\partial \vec X_{ab})^2\right)^2  }{|\vec X_{ab}|}\left(1+e^{-\frac{\tilde{\tilde{L}}|X|}{\tilde{\alpha}'}}\cdots \right) + \cdots\,.
\ea
Since $(2\pi\tilde{\tilde{L}})^2/(2\pi)^7\tilde{\tilde{g}}^2_s \tilde{\alpha}'^{4}$ is the tension of a single Kaluza-Klein monopole,
this moduli effective theory describes the moduli dynamics of the KK monopoles in IIA superstring theory.\footnote{
Under T-duality, the scalar field $X^9$ on NS5-branes corresponds to an additional isometry direction wound by KK monopoles,
and is mapped to the 0-form field on KK monopoles.
The gauge fields on NS5-branes and KK monopoles are the same.
(See Appendix \ref{app:fields}.)}
Written in the new parameters the scalar and thermal energy estimates (\ref{eq:NS5-thermal}) become
\ba
X_\text{KKM}^\text{(IIA)}\sim \frac{1}{\pi^2 N\tilde{\tilde{L}}T^2}
\qquad
E_\text{KKM}^\text{(IIA)}\sim \frac{1}{\pi^7 N \tilde{\tilde{g}}_s^2\tilde{\alpha}'^4 T^2}V_5\,.
\label{eq:KKM-thermal}
\ea
Note that the thermal energy $E_\text{KKM}^\text{(IIA)}$ does not depend on the radius $\tilde{\tilde{L}}$. It is reasonable that this direction is an isometry for KK monopoles.

\paragraph{Taub-NUT geometry from KK monopoles}
If the coupling in the type IIA KK monopole geometry is very strong, then the geometry lifts to the Taub-NUT geometry in M-theory. 
In this case we can check the that the thermal energy of the black KK monopole and the black Taub-NUT geometry agree
\ba
E_\text{KKM}^\text{(IIA)}\sim \frac{2\pi \tilde{\tilde{g}}_s \sqrt{\tilde{\alpha}'} V_5}{(2\pi)^8\tilde{\tilde{g}}_s^3\tilde{\alpha}'^{\frac92}NT^2}
=\frac{V_6}{(2\pi)^8l_\text{pl}^9 NT^2}
\sim E_\text{TN}
\ea
where we have used $V_6=2\pi \tilde{\tilde{g}}_s \sqrt{\tilde{\alpha}'} V_5$ and the relation (\ref{eq:M-relations}).
In this way, the 7d SYM theory describes the NS5-brane, KK monopole and Taub-NUT geometries through the moduli action. These results agree with the supergravity calculations \cite{Martinec:1999sa}.

%-----------------------------------------------------------------
\subsection{Notes: Relation of fields on M5/NS5/D5/KKM}
\label{app:fields}
%-----------------------------------------------------------------

So far we have discussed the connections between moduli theories of the various branes, focusing on the scalar moduli fields.
However these moduli theories also involve other bosonic fields.
Here we review the mapping of these fields in order to complete the connections of the moduli theories and the Yang-Mills description of the branes in string theory.

Let us list the bosonic fields with physical degrees of freedom on various branes:
\begin{itemize}
\item M5-brane: $B^{\rm (M5)}_{\mu\nu}, X_{\rm (M5)}^I$~~($\mu,\nu=0,\ldots,5$, $I=6,\ldots,10$)
\item IIA NS5-brane: $B^{\rm (NS5a)}_{\mu\nu}, C^{\rm (NS5a)}, X_{\rm (NS5a)}^I$~~($\mu,\nu=0,\ldots,5$, $I=6,\ldots,9$)
\item IIB NS5-brane: $A^{\rm (NS5b)}_\mu, X_{\rm (NS5b)}^I$~~($\mu=0,\ldots,5$, $I=6,\ldots,9$)
\item D5-brane: $A^{\rm (D5)}_\mu, X_{\rm (D5)}^I$~~($\mu=0,\ldots,5$, $I=6,\ldots,9$)
\item IIA KK monopole: $A^{\rm (KKa)}_{\mu}, C^{\rm (KKa)}, X_{\rm (KKa)}^I$~~($\mu,\nu=0,\ldots,5$, $I=7,\ldots,9$)
\item IIB KK monopole: $B^{\rm (KKb)}_{\mu\nu}, C^{\rm (KKb)}, \tilde C^{\rm (KKb)}, X_{\rm (KKb)}^I$~~($\mu,\nu=0,\ldots,5$, $I=7,\ldots,9$)
\end{itemize}
For KK monopoles, we set the 6th direction as the isometry direction.
All the $B_{\mu\nu}$'s are self-dual 2-form fields in 6d spacetime.
All the $C$'s and the $\tilde C$ are 0-form fields.
The relation among these fields are listed below~\cite{Eyras:1998hn}.
\begin{itemize}
\item 
The relation between the M5-brane and the IIA NS5-brane via the compactification of the M-theory circle:
\ba
B^{\rm (M5)}_{\mu\nu} = B^{\rm (NS5a)}_{\mu\nu}\,,\quad
X_{\rm (M5)}^I = X_{\rm (NS5a)}^I\,,\quad
X_{\rm (M5)}^{10} = C^{\rm (NS5a)}\,,
\ea
where $I=6,\ldots,9$. The M-theory circle is along the transverse space of NS5-brane.

\item
The relation between the IIA NS5-brane and the IIB NS5-brane via T-duality:
\ba
&&\back
B^{\rm (NS5a)}_{\mu\nu} = -\tilde A^{\rm (NS5b)}_{\mu\nu\tau}\,,\quad
B^{\rm (NS5a)}_{\mu\tau} = -A^{\rm (NS5b)}_\mu\,,\quad
C^{\rm (NS5a)} = -A^{\rm (NS5b)}_\tau\,,
\nt&&\back
X_{\rm (NS5a)}^I = X_{\rm (NS5b)}^I\,,
\ea
where $\mu,\nu\neq\tau$, and the T-duality direction $\tau$ is along the worldvolume of both NS5-branes.
$\tilde A^{\rm (NS5b)}_{\mu\nu\rho}$ is worldvolume dual of $A^{\rm (NS5b)}_\mu$.

\item 
The relation between the IIB NS5-brane and the D5-brane via S-duality:
\ba
A^{\rm (NS5b)}_\mu = A^{\rm (D5)}_\mu \,,\quad
X_{\rm (NS5b)}^I = X_{\rm (D5)}^I\,.
\ea

\item
The relation between the IIA KK monopole and the IIB KK monopole via T-duality:
\ba
&&\back
A^{\rm (KKa)}_\mu = B^{\rm (KKb)}_{\mu\tau}\,,\quad
A^{\rm (KKa)}_\tau = \tilde C^{\rm (KKb)}\,,\quad
\tilde A^{\rm (KKa)}_{\mu\nu\tau} = B^{\rm (KKb)}_{\mu\nu}\,,\quad
C^{\rm (KKa)} = C^{\rm (KKb)}\,,
\nt&&\back
X_{\rm (KKa)}^I = X_{\rm (KKb)}^I\,,
\ea
where $\mu,\nu\neq\tau$, and the T-duality direction $\tau$ is along the worldvolume of both KK monopoles.
$\tilde A^{\rm (KKa)}_{\mu\nu\rho}$ is worldvolume dual of $A^{\rm (KKa)}_\mu$.

\item
The relation between the IIA NS5-brane and the IIB KK monopole via T-duality:
\ba
&&\back
B^{\rm (NS5a)}_{\mu\nu} = -B^{\rm (KKb)}_{\mu\nu}+4\pi\alpha'Z^{\rm (KKb)}\partial_{[\mu} C^{\rm (KKb)} \partial_{\nu]}\tilde C^{\rm (KKb)}\,,\quad 
C_{\rm (NS5a)} = -\tilde C^{\rm (KKb)}\,,
\nt&&\back
X_{\rm (NS5a)}^6 = 2\pi\alpha'C^{\rm (KKa)}\,,\quad
X_{\rm (NS5a)}^I = X_{\rm (KKb)}^I\,,
\ea
where $I=7,\ldots,9$. 
The T-duality direction is along the transverse space of the NS5-brane and the Taub-NUT direction of KK monopole.
$Z^{\rm (KKb)}$ is the Taub-NUT coordinate of KK monopole.

\item
The relation between the IIB NS5-brane and the IIA KK monopole via T-duality:
\ba
A^{\rm (NS5b)}_\mu = -A^{\rm (KKa)}_{\mu}\,,\quad
X_{\rm (NS5b)}^6 = 2\pi\alpha' C^{\rm (KKa)}\,,\quad
X_{\rm (NS5b)}^I = X_{\rm (KKa)}^I\,,
\ea
where $I=7,\ldots,9$.
The T-duality direction is along the transverse space of the NS5-brane and the Taub-NUT direction of KK monopole.
\end{itemize}

%-----------------------------------------------------------------
%
\section{Moduli one-loop action with a compact spatial circle}
\label{app:GL}
%
%-----------------------------------------------------------------

In this Appendix we will derive the one-loop moduli action when one of the SYM spatial directions is compact, with radius $L$, and all fields are taken to be periodic on this. Note that this computation is different (and simpler) than the case discussed in \cite{Wiseman:2013cda} where Euclidean time is compact, but with anti periodic boundary conditions for the fermions.

Using 10-dimensional notation we may write our maximally supersymmetric Yang-Mills as,
\begin{eqnarray}
S = \frac{N}{\lambda} \int dt d^px \mathrm{Tr} \left[ - \frac{1}{4} F_{MN} F^{MN} - \frac{1}{2} \bar{\Psi} \gamma^M D_M  \Psi \right]
\end{eqnarray}
with $F_{MN} = \partial_M A_N - \partial_N A_M - i \left[ A_M, A_N \right]$ and with $\gamma^M$ in the appropriate Majorana-Weyl representation of SO(1,9). 
Consider a classical solution of the bosonic equations of motion $A_M = B_M$ where $B_M$ is diagonal. Then consider a gauge field constructed by expanding in fluctuations about this diagonal classical solution, $A_M = B_M + X_M$, where $X_M$ is the off-diagonal fluctuation. Adding ghosts and an appropriate gauge fixing term one obtains an action up to quadratic order in fluctuations $X_M$ and the fermions $\Psi$,
\begin{eqnarray}
S =  \frac{N}{\lambda} \int dt d^px \mathrm{Tr} \left[ L_0 + L_{2,X} + L_{2,\Psi} + L_{g.f.} + L_{ghosts} + \ldots \right]
\end{eqnarray}
where the leading term is,
\begin{eqnarray}
L_0 =  - \frac{1}{4} \bar{F}_{MN} \bar{F}^{MN}
\end{eqnarray}
with $\bar{F}_{MN} = \partial_A B_B - \partial_B B_A$, and the quadratic terms are $L_{2,X}$ for the fluctuations $X^M$ and $L_{2,\Psi}$ for the fermions, the gauge fixing term $L_{g.f.}$ which is quadratic in $X^M$, and $L_{ghosts}$ which is quadratic in the ghosts. There is no linear term in the fluctuation since $B_M$ is a classical solution. We choose to add the gauge fixing term $L_{g.f.} = \frac{1}{2} (\bar{D}_A X_A)^2$ where $\bar{D}_A = \partial_A - i \left[ B_A , \cdot \right]$. Then at quadratic order in $X_M$ one obtains,
\begin{eqnarray}
L^X_{quad} = L_{2,X} + L_{g.f.} = - \frac{1}{2} X_A \left[ - \eta_{AB} \bar{D}^C \bar{D}_C + 2 i \left[ \bar{F}_{AB} , \cdot \right] \right] X_B 
\end{eqnarray}
and for the fermions,
\begin{eqnarray}
L^\Psi_{quad} = L_{2,\Psi} = - \frac{1}{2} \bar{\Psi} \gamma^M \bar{D}_M  \Psi \,.
\end{eqnarray}
The leading term $L_0$ yields the classical moduli space action. Integrating out the quadratic fluctuations in $X_M$, the fermions $\Psi$, and the ghosts, then gives the one-loop correction to this moduli action.

Writing the diagonal matrix with colour indices $(B^M)_{ab} = a^M_a \delta_{ab}$ in terms of the diagonal moduli fields $a^M_a = ( a^\mu_a , \vec{\phi}_a )$, then we find the quadratic action,
\begin{eqnarray}
L^X_{quad} = - \frac{1}{2} \sum_{a \ne b}  {X_A}_{ab} \left[ - \eta_{AB} D^2_{ba} + 2 i {F}^{ba}_{AB} \right] {X_B}_{ba}  \; , \quad 
L^\Psi_{quad} = - \frac{1}{2} \sum_{a \ne b}  \bar{\Psi}_{ab} \gamma^M D^{ba}_M  \Psi_{ba}
\end{eqnarray}
where,
\begin{eqnarray}
D^{ba}_M = \partial_M - i \left( a^M_b - a^M_a \right) \; , \quad F^{ba}_{MN} = F^b_{MN} - F^a_{MN} \; , \; \mathrm{and} \quad F^a_{MN} = \partial_M {a_N}_a -  \partial_N {a_M}_a \,.
\end{eqnarray}
Integration over the hermitian off-diagonal boson matrices ${X_M}$ and ghosts yields the determinant,
\begin{align}
e^{i \, \Gamma_{B}} = \prod_{a<b} {\det} \left[ \triangle_{ba} \right]^2 \det \left[ \mathbb{\eta}_{MN} \triangle_{ba} + 2 i F^{ba}_{MN}  \right]^{-1} 
\end{align}
where,
\begin{align}
\triangle_{ba} = - {D^C}_{ba} {D_C}_{ba}=  - \left( \partial_\mu - i {a_\mu}_{ba} \right)^2 + | \vec{\phi}_{ba} |^2
\end{align}
returning now to the $(p+1)$-notation in the form of $\triangle_{ba}$, where $a^\mu_{ab} = a^\mu_a - a^\mu_b$ and similarly $\vec{\phi}_{ab} = \vec{\phi}_a - \vec{\phi}_b$. Integration over the fermions $\Psi$ gives the contribution,
\begin{align}
 e^{i\, \Gamma_{F}} = \prod_{a<b} \det \left[ \mathbb{I} \triangle_{ba} - \frac{i}{4} \left[ \gamma^M, \gamma^N \right] F_{MN}^{ba}\right]^{1/2} .
\end{align}

Firstly, we may evaluate the spacetime/spinor index dependence of these determinants as an expansion in the background field strength. We use the results,
\begin{align}
\det \left[ \mathbb{\eta}_{MN} \triangle + 2 i {F_{MN}} \right] & = \triangle^{10} + \triangle^8 \left( 2 F_2 \right) + \triangle^6 \left( -4 F_4 + 2 (F_2)^2 \right) + O\left( F^6 \right) 
\end{align}
where $F_2 = F_{MN} F^{MN}$ and $F_4 = F^{M}_{~N} F^{N}_{~O} F^{O}_{~P} F^{P}_{~M}$, 
and likewise,
\begin{align}
\det \left[ \mathbb{I} \triangle - \frac{i}{4} \left[ \gamma^M, \gamma^N \right] {F_{MN}} \right]^{1/2} &= \triangle^8 + \triangle^6 \left( 2 F_2 \right) + \triangle^4 \left( 2 F_4 + \frac{1}{2} (F_2)^2 \right) + O\left( F^6 \right) 
\end{align}
and then we see to leading order in the field strength that,
\begin{align}
\triangle^2 {\det} \left[ \mathbb{\eta}_{MN} \triangle + 2 i {F_{MN}} \right]^{-1} {\det}
&\left[ \mathbb{I} \triangle - \frac{i}{4} \left[ \gamma^M, \gamma^N \right] {F_{MN}} \right]^{\frac{1}{2}}  \nonumber \\
 & = \frac{  \triangle^8 + \triangle^6 \left( 2 F_2 \right) + \triangle^4 \left( 2 F_4 + \frac{1}{2} (F_2)^2 \right) + \ldots   }{ \triangle^{8} + \triangle^6 \left( 2 F_2 \right) + \triangle^4 \left( -4 F_4 + 2 (F_2)^2 \right) + \ldots  } \nonumber \\
 & =  \frac{ 1}{ \triangle^4 } \left( \triangle^4 - 6 \left( \frac{1}{4} (F_2)^2 - F_4  \right) + O( F^6 ) \right) .
\end{align}
Thus we deduce that to leading order in the field strength, 
\begin{align}
\label{eq:1loop}
e^{i\, \Gamma_{eff}} & = e^{i\, \left( \Gamma_{B} + \Gamma_{F} \right) } \nonumber \\
& = \prod_{a<b} \left( \det  \triangle_{ba}^4 \right)^{-1}  \det \left[ \triangle_{ba}^4 - 6 \left(  \frac{1}{4} \left( F^{ab}_{MN} F^{ab}_{MN} \right)^2  -  F^{ab}_{MN} F^{ab}_{NO} F^{ab}_{OP} F^{ab}_{PM}  \right) \right] .
\end{align}

Now we must evaluate these functional determinants. We wish to consider the case of our $(p+1)$-dimensional SYM where the $x^p$ spatial direction is compact with radius $L$, and all fields are periodic on this circle. 
Firstly consider the (real time) functional determinant,
\begin{align}
e^{i \, \Gamma} = {\det} \left[ \triangle + M \right] \; , \qquad \Delta = - \left( \partial_\mu - i a_\mu \right)^2 + \phi^2  
\end{align}
for a constant $M$, 
with $a^\mu$ and $\phi$ constant, and $\mu = 0, 1, \ldots , d-1$. 
Consider the heat equation,
\begin{align}
\partial_\sigma U(\sigma; x, x') = - \left(  \triangle_x + M \right) U(\sigma; x, x') \,.
\end{align}
Taking the boundary condition that $U(0; x, x') = \delta(x - x')$, then $U(\sigma; x, x')$ is the heat kernel, and we may represent the one-loop action due to this determinant as an appropriate trace of this heat kernel,
\begin{align}
\Gamma = - \int dt  d^{d-1} x \int_0^\infty \frac{d\sigma}{\sigma} U(\sigma; x, x) \,.
\end{align}
For constant $a^\mu$, $\phi$ and $M$ then the solution to this heat kernel is,
\begin{align}
U(\sigma; x, x') =  \frac{1}{\left( 4 \pi \sigma \right)^{d/2}} e^{- \sigma \left( \phi^2 + M \right) } e^{- \frac{1}{4 \sigma} | x - x' |^2 + i a^\mu \left( x - x' \right)_\mu } \,.
\end{align}
Hence we may use this heat kernel to represent our determinant as,
\begin{align}
\Gamma & = - \int d t \, d^{d-1} x \int_0^\infty d\sigma \frac{1}{\sqrt{ (4 \pi)^d \sigma^{d+2} } } e^{- \sigma \left( \phi^2 + M \right)} \end{align}
and we may use this to compute,
\begin{align}
e^{i \, \Gamma'} & = {\det} \left[ \triangle^4 - M^4 \right] =  {\det} \left[ \triangle + M \right]  {\det} \left[ \triangle - M \right]   {\det} \left[ \triangle + i M \right]   {\det} \left[ \triangle - i M \right] 
\end{align}
as,
\begin{align}
\Gamma' & = - \int d t \, d^{d-1} x \int_0^\infty d\sigma \frac{1}{\sqrt{ (4 \pi)^d \sigma^{d+2} } } e^{- \sigma \phi^2 }\left( e^{+ \sigma M} + e^{- \sigma M} + e^{i \sigma M} + e^{- i \sigma M} \right) . \nonumber \\
\end{align}
Now take $d = p+1$ and let us we have one compact spatial dimension, say $x^p$, with radius $L$, so $x^p \sim x^p + 2 \pi L$. Then we may split, $x^\mu = ( x^{\bar{\mu}}, x^p )$, where now $\bar{\mu} = 0, 1, \ldots p-1$. For a periodic field on the compact circle we may expand the determinant in Fourier modes as,
\begin{align}
{\det} \left[ \triangle + M \right] = \prod_{n \in \mathbb{Z}} {\det}' \left[ \bar{\triangle} + \left( \frac{n}{L} - a^p \right)^2 + M \right]
\end{align}
where now, $\bar{\Delta} = - \left( \partial_{\bar{\mu}} - i a_{\bar{\mu}} \right)^2 + \phi^2$
and ${\det}'$ signifies the determinant is taken over the continuous momenta of the non-compact $x^{\bar{\mu}}$ coordinates. Now we see that evaluating $e^{i \, \Gamma'} = {\det} \left[ \triangle^4 - M^4 \right]$ for this compact spatial circle yields,
\begin{align}
\label{eq:deteval}
\Gamma' & = - \sum_{n \in \mathbb{Z}} \int dt \, d^{p-1} x \int_0^\infty d\sigma \frac{1}{\sqrt{ (4 \pi)^p \sigma^{p+2} } }  e^{- \sigma \left( \phi^2 + \left( \frac{n}{L} - a^p \right)^2 \right) } \left( e^{+ \sigma M} + e^{- \sigma M} + e^{i \sigma M} + e^{- i \sigma M} \right) \nonumber \\
& \simeq - \sum_{n \in \mathbb{Z}} \int dt \, d^{p-1} x \int_0^\infty d\sigma \frac{1}{\sqrt{ (4 \pi)^p \sigma^{p+2} } } e^{- \sigma \left( \phi^2 + \left( \frac{n}{L} - a^p \right)^2 \right)  } \left( 4 + \frac{1}{6} \sigma^4 M^4 + O(M^6) \right)
\end{align}
to leading order in powers of $M$. Note that if we account for the fact that $a^\mu$ and $\phi$ are not constant we will obtain corrections which involve gradients of $a^\mu$ (via its field strength) and $\phi$, as,
\begin{align}
\Gamma' & \simeq - \sum_{n \in \mathbb{Z}} \int dt \, d^{p-1} x \int_0^\infty d\sigma  \frac{1}{\sqrt{ (4 \pi)^p \sigma^{p+2} } } e^{- \sigma \left( \phi^2 + \left( \frac{n}{L} - a^p \right)^2 \right)  } \left( 4 + \frac{1}{6} \sigma^4 M^4 + O(M^6) \right) \left( 1 + O\left( F_{\mu\nu}^2 ,  (\partial \phi)^2 \right) \right) .
\end{align}
Now using this determinant we may evaluate our SYM one-loop contribution \eqref{eq:1loop} as,
\begin{align}
 \Gamma_{eff} & = 
- \sum_{a<b} \sum_{n \in \mathbb{Z}} \int dt \, d^{p-1} x \int_0^\infty d\sigma \frac{1}{\sqrt{ (4 \pi)^p \sigma^{p+2} } } e^{- \sigma  \left( \left| \vec{\phi}_{ba} \right|^2 + \left( \frac{n}{L} - a^p_{ba} \right)^2 \right)  } \nonumber \\
& \qquad \times \left(  \sigma^4 \left(  \frac{1}{4} \left( F^{ab}_{MN} F^{ab}_{MN} \right)^2 - F^{ab}_{MN} F^{ab}_{NO} F^{ab}_{OP} F^{ab}_{PM}  \right)  + O\left( (F^a_{MN})^6 \right) \right)
\end{align}
and now performing the $\sigma$ integral we obtain,
\begin{align}
\Gamma_{eff} & = -  \sum_{a<b} \sum_{n \in \mathbb{Z}} \int dt d^{p-1} x  \frac{ \Gamma\left( \frac{8 - p}{2} \right)}{( 4 \pi)^{\frac{p}{2}}} \frac{ \left( \frac{1}{4} \left( F^{ab}_{MN} F^{ab}_{MN} \right)^2 - F^{ab}_{MN} F^{ab}_{NO} F^{ab}_{OP} F^{ab}_{PM} \right) }{  \left( \left( \frac{ n}{L} -  a^p_{ba} \right)^2 + \left| \vec{\phi}_{ba} \right|^2 \right)^{\frac{8 - p}{2}} } + O\left( (F^a_{MN})^6 \right) .
\end{align}
Then reinstating the compact $x^p$ integral, writing the coefficient in terms of an $(9-p)$-sphere volume, and continuing to Euclidean time, so that $\tau = i t$ and the Euclidean one-loop contribution to the action is $\Gamma^E_{eff}$ is given as $\Gamma_{eff} = i  \Gamma^E_{eff}$, then,
\begin{align}
\Gamma^E_{eff} & = -  \sum_{a<b} \sum_{n \in \mathbb{Z}} \int d\tau d^{p-1} x \frac{1}{2 \pi L} \int_0^{2 \pi L} dx^p \frac{ (2 \pi)^{5-p}}{32 (8 - p) \Omega_{(9-p)}} \frac{  \left( 4 F^{ab}_{MN} F^{ab}_{NO} F^{ab}_{OP} F^{ab}_{PM}  -  \left( F^{ab}_{MN} F^{ab}_{MN} \right)^2 \right) }{  \left( \left( \frac{n}{L} -  a^p_{ba} \right)^2 + \left| \vec{\phi}_{ba} \right|^2 \right)^{\frac{8 - p}{2}} } 
\nn \\ & \qquad + O\left( (F^a_{MN})^6 \right)
\end{align}
giving the result \eqref{eq:int-compact} in the main text. 
We note that in reinstating the integral it appears we have neglected field dependence in the $x^p$ direction. However, to leading order in derivatives, including such dependence will not change the result. Since we are working in a gradient expansion, we have already assumed that field variation in the $x^p$ direction, or indeed any other direction, is controlled.

Let us briefly consider recovering the one-loop term for the de-compactified $L \to \infty$ limit in equation \eqref{eq:non-thermal}.
We note that we may write the infinite sum as,
\begin{align}
\sum_{n \in \mathbb{Z}} & \frac{\Gamma\left( \frac{8 - p}{2} \right)}{ \left( \left( \frac{ n}{L} - a \right)^2 + \left| \vec{\phi} \right|^2 \right)^{\frac{8 - p}{2} }} \nonumber \\
& =
\Gamma\left( \frac{7 - p}{2} \right) \sqrt{ \pi }L | \vec{\phi} | ^{- (7-p)} + \frac{2^{ - \frac{5 - p}{2} }(2\pi L)^{8 - p} }{\sqrt{\pi}} \sum_{n = 1}^{\infty}  \left( \frac{n}{2\pi L | \vec{\phi} |} \right)^{ \frac{7-p}{2} } K_{ - \frac{7-p}{2}}\left( 2\pi n L | \vec{\phi} | \right) \cos\left( 2\pi n L a \right)  \end{align}
and then for $L \left| \vec{\phi} \right| \to \infty$ we find,
\begin{align}
\sum_{n \in \mathbb{Z}}  \frac{\Gamma\left( \frac{8 - p}{2} \right)}{ \left( \left( \frac{ n}{L} - a \right)^2 + \left| \vec{\phi} \right|^2 \right)^{\frac{8 - p}{2} }}  \simeq
\Gamma\left( \frac{7 - p}{2} \right) \sqrt{ \pi }L | \vec{\phi} | ^{- (7-p)} 
 \end{align}
 plus exponentially small corrections, 
and thus in this de-compactified limit,
\begin{align}
\Gamma_{eff} & = -  \sum_{a<b}  \int dt d^{p-1} x  \frac{1}{2 \pi L} \int_0^{2 \pi L} dx^p \left( \frac{ \sqrt{\pi} L \Gamma\left( \frac{7 - p}{2} \right)}{( 4 \pi)^{\frac{p}{2}}} \right) \frac{ \left( \frac{1}{4} \left( F^{ab}_{MN} F^{ab}_{MN} \right)^2 - F^{ab}_{MN} F^{ab}_{NO} F^{ab}_{OP} F^{ab}_{PM} \right) }{  \left| \vec{\phi}_{ba} \right|^{7-p} }
\nn \\ & \qquad + O\left( (F^a_{MN})^6 \right)
\end{align}
and hence,
\begin{align}
\Gamma^E_{eff} & = -  \sum_{a<b}  \int d\tau d^{p-1} x  \int dx^p \frac{ (2 \pi)^{4-p} }{32 (7-p) \Omega_{(8-p)} } \frac{ \left(  4 F^{ab}_{MN} F^{ab}_{NO} F^{ab}_{OP} F^{ab}_{PM} -  \left( F^{ab}_{MN} F^{ab}_{MN} \right)^2 \right) }{  \left| \vec{\phi}_{ba} \right|^{7-p} }
\nn \\ & \qquad + O\left( (F^a_{MN})^6 \right)
\end{align}
which gives the result in equation \eqref{eq:non-thermal}.

\section{Moduli theory and the boosted Schwarzschild  black hole}
\label{app-D0}
As discussed in \ref{sec:D0}, 
a phase transition occurs in the $N$ D0-brane theory at low temperature, related to a GL transition of the boosted black string in eleven dimensions.
The low temperature phase is related to the  boosted Schwarzschild black hole which carries momentum $N/R$, localised on the M-theory circle of radius $R$.
A similar transition also occurs in the wave theory (\ref{eq:wave-moduli-example}) too.
In this Appendix, we review how the moduli theory explains the boosted Schwarzschild black hole.
We begin by reviewing the boosted Schwarzschild black hole in gravity in \ref{app:11DBH} and then turn to the moduli theory description in \ref{app:D0MM}.

\subsection{Boosted Schwarzschild black hole on a circle }
\label{app:11DBH}
Let us consider a $D$-dimensional Schwarzschild black hole boosted along a compact circle of radius $R$.
To begin, the $D$-dimensional Schwarzschild black hole solution is
\begin{align}
ds^2= -fdt^2 + \frac{1}{f}dr^2 +r^2 d \Omega_{D-2}^2, \qquad f=1-\left(\frac{r_0}{r}\right)^{D-3}
\end{align}
with mass $M$ and entropy $S$ given by
\begin{align}
M=\frac{\Omega_{D-2}}{2\kappa_{D}^2}(D-2)r_0^{D-3}, \qquad 
S=\frac{2\pi \Omega_{D-2}}{\kappa_{D}^2}r_0^{D-2}.
\end{align}
Here $\kappa^2_{D}$ is the $D$-dimensional gravitational coupling.
Now we boost this solution along the $D$-th direction with the boost parameter $\alpha$.
The energy $E$ and the $D$-th momentum $P$ become
$E=M \cosh \alpha$ and 
$P=M \sinh \alpha$, respectively.
The event horizon does not Lorentz contract \cite{Horowitz:1997fr} and the entropy does not change.
We compactify the $D$-th direction with the radius $R$, and take $\alpha$ large by keeping $E_{\text{LC}}=E-P$ finite, where $P=N/R$.
This condition fixes $e^\alpha = 2N/RM$ and $E_{\text{LC}}=RM^2/2N$.
Then the temperature is obtained by $T=dE_{\text{LC}}/dS$, and we can write down the physical quantities as functions of temperature,
\begin{align}
\label{eq:D-SSBH}
r_0&=\left(
\frac{8 \pi N \kappa_{D}^2 T}{(D-2)(D-3)\Omega_{D-2} R } \right)^{\frac{1}{D-4}}, 
\quad
E_{\text{LC}}= \left( \frac{\kappa^2_{D}}{(D-2) \Omega_{D-2}} \right)^{\frac{2}{D-4}}
\left( \frac{8N}{R} \right)^{\frac{D-2}{D-4}} \left(\frac{\pi T}{D-3}\right)^{\frac{2(D-3)}{D-4}}, \nonumber \\
S&=(2 \pi)^{\frac{2(D-3)}{D-4}} 
\left( \frac{\kappa^2_{D}}{ \Omega_{D-2}} \right)^{\frac{2}{D-4}}
\left( \frac{4NT}{(D-3)(D-2)R} \right)^{\frac{D-2}{D-4}}.
\end{align}
Note that these expressions are valid when the size of the horizon $r_0$  is much smaller than the size of the boosted compact circle $e^{\alpha} R$.
In the case where the horizon size is comparable to the circle size, finite volume effects cannot be ignored.
In particular, a Gregory-Laflamme transition to a black string occurs
when $r_0 \sim e^{\alpha} R$. This implies that the transition occurs around $S \sim N$.
Thus an expression for the critical temperature of the transition is given by,
\begin{align}
T_c \sim \frac{R}{(\kappa_D^4 N^2)^{\frac{1}{D-2}}}
\label{eq:T_c-D-dBH}.
\end{align}
Beyond this temperature, the stable configuration is the boosted black string which is equivalent to the $D-1$-dimensional black 0-brane with $\mu=1/R$ and $\kappa^2= \kappa_D^2/2\pi R$.
Its energy and the horizon radius are given as
\begin{align}
E= \frac{(D-2)\Omega_{D-3}}{4} \frac{2\pi R }{\kappa_D^2}z_0^{D-4}, \qquad
r_0=&\left(\frac{2^4 \pi NT^2 \kappa_D^2}{(D-4)^3 \Omega_{D-3}R^2} \right)^{\frac{1}{D-6}}
\label{eq:D0BH-general}
\end{align}

\subsection{Boosted Schwarzschild black hole  from moduli theory}
\label{app:D0MM}

We now argue how the $D$-dimensional boosted Schwarzschild black hole in the previous subsection can be explained using the moduli theory \cite{Morita:2013wfa} for the 0-brane in $D-1$ dimensions,
\begin{align}
S_{\text{moduli}}=& \int d \tau  \frac{1}{2}  \sum_{a=1}^N \frac{1}{R}  \left(\partial_\tau X_a\right)^2 -
 \int d \tau  \sum_{a,b=1}^N  \frac{\kappa_D^2}{8 \pi R (D-4) \Omega_{D-3}}  \frac{ \left( \frac{1}{R} (\partial_\tau X_{ab})^2 \right)^2  }{X_{ab}^{D-4}} + \cdots.
\label{eq:D0-general-action}
\end{align}
Here we have taken the Newton constant $\kappa_{D-1}^2 = \kappa_D^2/ 2 \pi R $ and the mass of the 0-brane $\mu =1/R$ by noting that the 0-branes are obtained from the dimensional reduction of the $D$-dimensional wave.

As discussed in Section \ref{sec:D0}, Li and Martinec\cite{Li:1998ci} considered the contribution of the zero-energy bound state in the 0-brane system.
They argued that if the 0-branes form $M$ zero-energy bound states, the effective action (\ref{eq:D0-general-action}) gets modified to (from now on we omit  numerical factors, except factors of $\pi$) 
\begin{align}
S_{\text{moduli}}^{\text{bound}}=& \int d \tau \sum_{a=1}^M m_{\text{bd}}   \left(\partial_\tau X_a\right)^2 -
 \int d \tau  \sum_{\delta p_-=0}^{N/M} \sum_{a,b=1}^M  f_D(\delta p_-) \frac{\kappa_D^2}{ \pi R  \Omega_{D-3}}  \frac{ \left( m_{\text{bd}} (\partial_\tau X_{ab})^2 \right)^2  }{X_{ab}^{D-4}} + \cdots.
\label{one-loop-D0-low}
\end{align}
Here the 0-branes in each bound state is $N/M$ and the mass of the bound state is equal to $m_{\text{bd}} \equiv  \frac{N}{M} \frac{1}{R}$.
$f_D(p_-)$ symbolically represents the effect of the exchanges of the longitudinal momentum in the four point interaction of the bound states.

\cite{Li:1998ci} found that if $M$ is assumed to be the same order as the entropy of the system, then the boosted black hole thermodynamics is reproduced from the effective action (\ref{one-loop-D0-low}) with an assumption $f(p_-)\sim 1$.
We review this derivation in a slightly different way\footnote{
 Li and Martinec's derivation\cite{Li:1998ci} and ours are slightly different although the results are same.
They used three assumptions: (I) the saturation of the uncertainty relation $ p_{\text{bd}}X \sim m_{\text{bd}} (\partial_\tau X)X \sim 1$, (II) the number of bound states $M$ in Eq.(\ref{one-loop-D0-low}) is the same order as the entropy of the system, $M \sim S_{\text{entropy}}$ and (III) the amplitude satisfies $f(p_-) \sim O(1)$.
In our derivation, instead of the assumption (I), we assume $  \partial_\tau X \sim \pi T X$, since we know that this assumption gives correct results for the black branes discussed in Section \ref{sec:SYM}.
In addition,\cite{Li:1998ci} did not evaluate the thermodynamics of the IIA D0-brane black hole at $T>T_c $ where their assumption (I) does not work.
} based on our moduli effective action proposal.
We assume $ \partial X \sim \pi TX$ and estimate the classical and the interaction terms of the effective action as
\begin{align}
S_{\text{classical}} \sim  \frac{N}{R} \pi^2 T X^2
, 
\qquad S_{\text{one-loop}} \sim \frac{N}{M} \left(\frac{f_D  \kappa_D^2}{\pi \Omega_{D-3}}\right) \frac{\pi^4 N^2 T^3}{R^3 X^{D-8}} 
.
\label{evaluate-11dimBH}
\end{align}
Here we have assumed $f_D(p_-)=f_D$ is a constant.
Then we obtain 
\begin{align}
X \sim \left( \left(\frac{f_D \kappa_D^2}{\pi \Omega_{D-3}}\right)  \frac{N^2 \pi^2 T^2 }{M R^2} \right)^{\frac{1}{D-6}}
\end{align}
through the virial theorem, and the total energy is estimated as
\begin{align}
 E \sim
 \left(\frac{f_D \kappa_D^2}{\pi \Omega_{D-3}}\right)^{\frac{2}{D-6}} 
\left( \frac{N}{R} \right)^{\frac{D-2}{D-6}} \frac{ \left(\pi T \right)^{\frac{2(D-4)}{D-6}}
}{M^{\frac{2}{D-6}}}\,.
\end{align}
Now we assume 
\begin{align}
S_{\text{entropy}}\sim \pi M,
\label{eq:assumption-Li}
\end{align}
just as \cite{Li:1998ci}  assumed that the number of bound states is the same order as the entropy.
The $\pi$ dependence in this assumption will be explained later.
Then we substitute an ansatz $ E = C S_{\text{entropy}}^x$ and $T= \partial E/ \partial S_{\text{entropy}}= x C S_{\text{entropy}}^{x-1} $ into this equation, and we can solve for $x$ and $C$ as
\begin{align}
x=\frac{2(D-3)}{D-2}, \qquad C \sim \frac{R}{N}  \left(\frac{f_D \kappa_D^2}{\pi \Omega_{D-3}}\right)^{-\frac{2}{D-2}} \pi^{-\frac{2(D-3)}{D-2}}.
\end{align}
Thus we obtain
\begin{align}
 E & \sim
 \frac{R}{N}  \left(\frac{f_D \kappa_D^2}{\pi \Omega_{D-3}}\right)^{-\frac{2}{D-2}} \pi^{-\frac{2(D-3)}{D-2}} S^{\frac{2(D-3)}{D-2}} \sim
 \left(  \frac{N}{R}\right)^{\frac{D-2}{D-4}}  \left(\frac{f_D \kappa_D^2}{\pi \Omega_{D-3}}\right)^{\frac{2}{D-4}}  \left(\pi T \right)^{\frac{2(D-3)}{D-4}}, \nt
S_{\text{entropy}} & \sim \pi^{\frac{2(D-3)}{D-4}} \left(  \frac{f_D \kappa_D^2}{\pi \Omega_{D-3}}    \right)^{\frac{2}{D-4}} \left(\frac{NT}{R} \right)^{\frac{D-2}{D-4}}, \nt
X & \sim \left( \left(\frac{f_D \kappa_D^2}{\pi \Omega_{D-3}}\right) \frac{N  \pi T }{ R }\right)^{\frac{1}{D-4}}.
\label{eq:0-brane-bound-result}
\end{align}
If we assume $f_D \sim O(1)$,
these results are consistent with the thermodynamics of the boosted Schwarzschild  black hole (\ref{eq:D-SSBH}) except the $\pi$ dependence.
To reproduce the $\pi$ dependence, we need to impose
\begin{align}
f_D \sim \frac{\pi \Omega_{D-3}}{\Omega_{D-2}}.
\end{align}

What do these results mean?
Although we have employed several assumptions, since we reproduce the supergravity predictions  we do capture the correct nature of the matrix quantum mechanics at low temperature.
Li and Martinec pointed out that $ p_{\text{bd}} X \sim m_{\text{bd}} (\partial_\tau X) X \sim 1$ is satisfied at the localised black hole phase, where $p_{\text{bd}}$ is the momentum of the bound state, 
and the uncertainty relation seems saturated.
Presumably the amplitude $f(p_-)$ somehow becomes $O(1)$ when the uncertainty relation is saturated\footnote{If we assume that $f(p_-) \sim O(1)$ is satisfied at arbitrary number of the bound states $M$, we may obtain a configuration which has a larger entropy and it will be favoured thermodynamically.
Since this result is inconsistent with the supergravity prediction, $f(p_-) \sim O(1)$ may be not always satisfied. }.
This feature might be related to the instability of the $D-1$ dimensional black 0-brane (\ref{eq:D0BH-general}).
Indeed in this configuration, the single 0-brane satisfies
\begin{align}
p X \sim \frac{1}{R} \left(\partial_\tau X\right) X \sim \left( \frac{T}{T_c} \right)^{\frac{D-2}{D-6}} <1 , \qquad (T < T_c, D>6)
 \end{align}
 where $T_c$ is given by (\ref{eq:T_c-D-dBH})
 and the uncertainty relation seems violated\footnote{$D=6$ is the Hagedorn case. In $D=5$, the black holes have negative specific heat and the 4 dimensional black 0-brane is stable at $T< T_c$. Again the uncertainty relation is broken when the black 0-brane is unstable at $T> T_c$.}.
  
Finally we comment on the $\pi$ dependence of the calculations in this section.
In the calculations, we have used three assumptions: $\partial_\tau X \sim \pi T X$, 
$S_{\text{entropy}} \sim \pi M$ and $f_D \sim \pi \Omega_{D-3}/\Omega_{D-2}$.
We do not have any natural reason to fix the $\pi$ dependences in these equations.
The first assumption $\partial_\tau X \sim \pi T X$ is consistent with various branes as argued in Section \ref{sec:Dp-brane-gravity}, and it may be natural to impose the same $\pi$ dependence in the current case too.
Concerning $S_{\text{entropy}} \sim \pi M$, we fixed the $\pi$ dependence so that the saturation of the uncertainty relation  $ p_{\text{bd}} X \sim m_{\text{bd}} (\partial_\tau X) X \sim 1$ is satisfied, including the $\pi$ dependence.
(We should emphasise that there is no reason that we expect the saturation of the uncertainty relation including the $\pi$ dependence.)
Regardless of what $f_D$ is, the saturation of the uncertainty relation follows from the first two assumptions.
The last assumption $f_D \sim \pi \Omega_{D-3}/\Omega_{D-2}$, would need to be confirmed, for instance by calculating the scattering of bound states.
We leave this challenging issue for future work.

%-----------------------------------------------------------------
%
\bibliographystyle{JHEP}
\bibliography{dpbranebibV5}
%
%-----------------------------------------------------------------

\end{document}